\documentclass[12pt,a4paper]{article}
\usepackage[T1]{fontenc}

\usepackage{amsmath,amssymb,slashed,cancel}
\usepackage{cite,tabularx}
\usepackage{subcaption}
\usepackage{graphicx,color}
\usepackage[normalem]{ulem}
\usepackage{mathtools}
\usepackage{enumerate}
\usepackage{ascmac}
\usepackage{enumitem}
\usepackage{float}
\usepackage{placeins}
\usepackage{multirow}
\allowdisplaybreaks

\usepackage[height=8.85in,width=6.4in]{geometry}
\renewcommand{\baselinestretch}{1.1}
\usepackage{comment}
\newcommand\package[2][\relax]{\texttt{#2\ifx#1\relax\relax\relax\else\,\linebreak[0]#1\fi}}

\numberwithin{equation}{section} 
 
\def\beq#1\eeq{\begin{align}#1\end{align}}

\definecolor{BlueViolet}{rgb}{0.2, 0.00, 0.7}
\definecolor{Blue}{rgb}{0.15, 0.00, 0.9}
\usepackage[colorlinks=true,linkcolor=Blue,citecolor=Blue,urlcolor=BlueViolet]{hyperref}

\begin{document}
\begin{titlepage}
\setcounter{page}{0} 

\begin{center}

\vskip .55in

\begingroup
\centering
\large\bf 
Muon collider experiments as electron/positron beam sources:
case studies of new light-particle searches
\endgroup

\vskip .4in

{
Yasuhito Sakaki$^{\rm (a)}$ and  Daiki Ueda$^{\rm (b)}$
}

\vskip 0.4in

\begingroup\small
\begin{minipage}[t]{0.9\textwidth}
\centering\renewcommand{\arraystretch}{0.9}
\begin{tabular}{c@{\,}l}
$^{\rm(a)}$
&High Energy Accelerator Research Organization (KEK), Tsukuba\\
&Ibaraki 305-0801, Japan
\\
$^{\rm(b)}$
& Physics Department, Technion \text{--} Israel Institute of Technology,
Technion city\\
&Haifa 3200003, Israel\\
\end{tabular}
\end{minipage}
\endgroup

\end{center}

\vskip .4in

\begin{abstract}\noindent
At muon colliders, muon decays naturally produce intense electrons and positrons with unique features, namely high energies, high repetition rates, and small intrinsic uncertainties, that are unavailable at existing accelerator facilities.
We quantitatively study the feasibility of extracting such particles in two representative future muon collider designs, IMCC and $\mu$TRISTAN.
Using Monte Carlo simulations with the corresponding design parameters, we study the spatial, angular, and energy distributions of decay electrons and positrons in the curved sections of the collider ring.
We find that typical deflections of $0.1$\text{--}$10~\mathrm{mrad}$ can be achieved even for high-energy electrons carrying large energy fractions ($\simeq 0.6$\text{--}$1.0$) of the muon beam energy, with the ring bending magnets (or magnets providing an equivalent field) effectively serving as a  \lq\lq pre-septum magnet\rq\rq\, that partially deflects the beam before the main septum magnet, suggesting that the extraction scheme could be practically feasible.
Exploiting the distinct beam properties of IMCC and 
$\mu$TRISTAN, we propose complementary search strategies --- missing energy and momentum searches for dark matter at 
$\mu$TRISTAN and visible-decay searches for axion-like particles and light scalars at IMCC --- which probe parameter space beyond the reach of current and other proposed experiments.

\end{abstract}
\end{titlepage}

\setcounter{page}{1}
\renewcommand{\thefootnote}{\#\arabic{footnote}}
\setcounter{footnote}{0}

\begingroup
\renewcommand{\baselinestretch}{1} 
\setlength{\parskip}{2pt}          
\hrule
\tableofcontents
\vskip .2in
\hrule
\vskip .4in
\endgroup

\section{Introduction}\label{sec:intro}

Muons provide a unique opportunity to realize a compact high-energy collider experiment due to their status as elementary particles with large mass.
Muon colliders planned with center-of-mass energies of $\mathcal{O}(1\text{--}10)$ TeV~\cite{Delahaye:2019omf,Bartosik:2020xwr,Long:2020wfp,MuonCollider:2022ded,MuonCollider:2022nsa,Accettura:2023ked,InternationalMuonCollider:2024jyv,InternationalMuonCollider:2025sys,Hamada:2022mua} offer promising prospects for discovering new heavy particles, extending the mass sensitivity well beyond the capabilities of the LHC.
In addition, they will enable precise studies of the electroweak and Higgs sectors, benefiting from the abundant vector boson initiated processes at the electroweak scale and the suppressed QCD background.
Therefore, a high-energy muon collider is anticipated to serve as the next major step in probing fundamental physics beyond the HL-LHC, while complementing the physics program of future Higgs factory experiments.

However, muon decays generate neutrinos and electrons/positrons, which can induce beam-related backgrounds in the detector, cause magnet quenches, and result in long-term radiation damage that may lead to magnet failures.
In particular, the decay electrons and positrons can carry TeV-scale energies, depositing energy via electromagnetic showers in the surrounding material.
Thus, the design of muon collider experiments requires dedicated radiation shielding configurations.
To mitigate such radiation, absorber materials are planned to be installed to shield detectors, magnets, and other sensitive components~\cite{Mokhov:2011zza, Kashikhin:2012zza, Mokhov:2014lkf,Calzolari:2022uyb,Lechner:2024nol,InternationalMuonCollider:2024jyv,InternationalMuonCollider:2025sys}.
In contrast, neutrino radiation is actively investigated for its potential in neutrino physics~\cite{Kitano:2024kdv,Adhikary:2024tvl,Liu:2024ywd,Bojorquez-Lopez:2024bsr,Kling:2025zsb}.
The decay neutrinos can pass through the collider ring and could be utilized by installing additional far detectors.
These considerations naturally motivate exploring ways to leverage the decay electrons and positrons for physics studies.

The decay electrons and positrons produced in muon collider experiments possess two unique characteristics rarely found at conventional electron and positron accelerator facilities --- high energy, and the potential for continuous beams --- which make them promising for new physics applications.
First, as noted above, the decay electrons and positrons carry TeV-scale energies, as they originate from the decays of muons with TeV-scale energies in the laboratory frame.
Second, in contrast to most conventional accelerator electron and positron beams, the decay electrons and positrons can potentially form a continuous, unbunched beam, despite originating from bunched muon beams.
This is because the highly boosted muon beams decay with an approximately uniform probability along the entire collider ring.
In fixed-target experiments, bunched beams produce many processes simultaneously at the target, making it difficult to identify each beam-induced event separately.
Compared to bunched beams, continuous beams typically have a lower particle flux; however, they offer the advantage of identifying beam-induced processes on an event-by-event basis.
The DM searches based on missing momentum/energy require identifying beam-induced processes on an event-by-event basis, making continuous beams essential, as in experiments such as NA64~\cite{NA64:2016oww,Gninenko:2016kpg} and LDMX~\cite{Izaguirre:2014bca,Mans:2017vej}.

Examining the potential utilization of electron and positron beams with unique properties at muon colliders is a timely and valuable endeavor, particularly in light of ongoing discussions regarding the feasibility of future muon collider projects.
In this paper, we quantitatively analyze the feasibility of extracting electron and positron beams in muon collider experiments, drawing on the LHC beam extraction technology, in particular the kicker\text{-}septum system~\cite{Goddard:2003ni,Goddard:2003dr} and its demonstrated beam deflection performance.
At the LHC, kicker and septum magnets are employed to deflect and extract proton beams toward the beam dump, ensuring safe collider operation.
Similar beam extraction technology is also planned to be implemented at the FCC\text{-}hh~\cite{PhysRevAccelBeams.20.041002}.
Assuming the geometrical size and magnetic field strength of the bending magnets installed along the curved sections of a muon collider ring, we analyze the positions, angles, and energy distributions of electron and positron beams generated within a region of length $L_{\rm ext}$ along the beam line.
Because decay electrons inherit lower momenta than their parent muons, they experience larger bending in a given magnetic field.
This feature is advantageous, as the bending magnets (or magnets providing an equivalent magnetic field) in the ring may naturally function as a \lq\lq pre-septum magnet\rq\rq\, without the need for an additional kicker.
We show that deflections of $0.1$\text{--}$10~\mathrm{mrad}$ --- significantly larger than those in the LHC extraction system --- can typically be achieved even for high-energy electrons with large energy fractions $\simeq 0.6$\text{--}$1.0$ of the muon beam energy, suggesting that such an extraction scheme could be practically feasible and attractive.
Interestingly, the number of extractable electrons is independent of the extraction length $L_{\rm ext}$ as a consequence of beam-transport acceptance.
Further details are presented in Sec.~\ref{sec:cont}.

To highlight the potential of electron and positron beams, we propose searches for new light particles using them and evaluate the corresponding expected sensitivities.
As discussed below, two muon collider proposals, $\mu$TRISTAN~\cite{Hamada:2022mua} and the International Muon Collider Collaboration (IMCC)~\cite{InternationalMuonCollider:2024jyv,InternationalMuonCollider:2025sys}, feature quite different electron and positron beam properties.
Accordingly, we propose two complementary search strategies for light particles: a missing energy/momentum search in the $\mu$TRISTAN setup and a visible-decay search in the IMCC setup.
The proposed experimental setups for these searches are illustrated in Figs.~\ref{fig:LDMX} and \ref{fig:setup_ALP}.

In the missing energy/momentum search, our main target is dark matter (DM) particles in the MeV--GeV mass range.
A stable particle in this mass range interacting with the SM via a hidden-photon mediator constitutes a well-motivated DM candidate.
For suitable model parameters, its thermal relic density can account for the observed DM abundance while remaining consistent with current astrophysical and experimental constraints.
In our proposal, DM particles are produced in a thin fixed target (with a thickness much smaller than its radiation length) through hidden-photon exchange in interactions between the continuous electron and positron beams and the target.
Tagger and recoil trackers placed upstream and downstream of the fixed target are designed to measure the momenta of SM particles and to reconstruct missing-momentum signatures indicative of DM production.
In the visible-decay search, our main targets are mediator particles between the SM and dark sectors, including axion-like particles and light scalars coupled to photons.
These new particles can be produced in a thick fixed target (with a thickness much larger than its radiation length) through interactions of the bunched electron and positron beams with the target.
An electromagnetic calorimeter placed downstream of the fixed target is designed to measure energy deposition from decays of these particles into SM final states, such as photons.
Together, these search strategies based on extracted electron and positron beams can be pursued in parallel with the primary physics program at the main interaction point, thereby significantly broadening the muon collider's capability to search for physics beyond the SM.

The rest of the paper is organized as follows.
In Section~\ref{sec:cont}, we analyze the positions, angles, and energy distributions of electron and positron beams produced by muon beam decays in the curved section of the muon collider ring, including the effects of the magnetic fields.
The analysis is performed using the Monte Carlo simulation tool PHITS~\cite{Sato:2023qsv}.
In Section~\ref{sec:DM}, based on the electron and positron beams described in Section~\ref{sec:cont}, we assess the potential for searches for new light particles.
We consider two benchmark scenarios: dark matter mediated by a dark photon, and ALPs and light scalars coupled to photons.
For each case, we present sensitivity estimates for fixed-target experiments using the extracted electron and positron beams.
Section~\ref{sec:sum} summarizes the main findings of our analysis.
Appendix~\ref{app:Irre} provides additional details on our estimates of irreducible backgrounds in the missing energy/momentum search.

\section{Electron and positron beams at muon collider experiments}
\label{sec:cont}

Muon collider experiments produce high-repetition-rate electrons and positrons originating from muon decays in the collider ring, and these particles are essentially free of hadronic contamination.
Although these particles have mainly been studied in the context of radiation shielding for accelerator components and detector systems~\cite{InternationalMuonCollider:2024jyv,InternationalMuonCollider:2025sys}, their sufficiently controlled extraction as a beam could also be exploited, for example, in new physics searches.
In this section, we use the parameter sets listed in Table~\ref{tab:para}, proposed by the International Muon Collider Collaboration (IMCC) and $\mu$TRISTAN, as benchmarks and simulate the corresponding electron-beam fluxes in the magnetic fields of the collider rings with PHITS~\cite{Sato:2023qsv}.
Although the results presented below are equally valid for positrons produced in $\mu^+$ decays, we restrict our discussion to electrons from $\mu^-$ decays for clarity throughout this paper.

\begin{table}[t]
    \renewcommand{\arraystretch}{1.2}
	\begin{center}
		\begin{tabular}{|c|c|c|c|c|c|} \hline\hline 
			Parameter  & IMCC-1-I  & IMCC-1-II & IMCC-2 & $\mu$TRISTAN 
			\\ \hline
			Muon energy (lab frame) & 1.5 TeV & 5 TeV & 5 TeV & 1 TeV
			\\                            
			 Collider circumference $(C_{\rm coll})$ & $4.5$ km & 10 km & 15 km & 3 km 
			\\  
			Muons/bunch & $2.2\times 10^{12}$ & $1.8\times 10^{12}$ & $1.8\times 10^{12}$ & $2.2\times 10^{10}$
			\\            
			Bunches/train & 1 & 1 & 1 & 20(40)
			\\            
			Repetition rate & 5 Hz & 5 Hz & 5 Hz & 50 Hz
			\\
            \hline
			$\hat{N}_{e^\pm} ~ {\rm [1/year]}$ (Eq.~\eqref{eq:hatNe}) & 
            $2.9\times10^{14}$ & $3.6\times10^{14}$ & $2.4\times10^{14}$ & $5.9(12)\times10^{14}$
			\\
			Repetition rate of $e^\pm$ (Eq.~\eqref{eq:rep_def}) & 67~kHz & 30~kHz & 20 kHz & 2(4)~MHz
			\\
			$\hat{n}_{e^\pm/{\rm bunch}}$ (Eq.~\eqref{eq:hatnperbunch}) & 2400 & 1900 & 1900 & 24
			\\
			\hline  \hline  
		\end{tabular}
\caption{Tentative parameters presented by the International Muon Collider Collaboration (IMCC)~\cite{InternationalMuonCollider:2024jyv,InternationalMuonCollider:2025sys} for three cases and for $\mu$TRISTAN~\cite{Hamada:2022mua}.
In the IMCC, cardinal numbers (1 and 2) refer to Scenarios 1 and 2
(Energy Staging and Luminosity Staging, respectively),
while Roman numerals (I and II) indicate the two stages.
For Scenario 2, the stages are not distinguished, as they yield no
significant difference in the analysis below.
Throughout this paper, one year of operation denotes $1.2\times 10^7$ sec.
Before accounting for the efficiency factor~\eqref{eq:eta}, the expected number of extractable electrons over one year of operation $\hat{N}_{e^\pm}$ is largely independent of the experimental configuration.
In contrast, the IMCC provides a much larger number of electrons per muon bunch $\hat{n}_{e^\pm/{\rm bunch}}$, while $\mu$TRISTAN offers a significantly higher repetition rate of $e^\pm$.
These differences suggest that the optimal applications of the extracted electron/positron beams vary depending on the muon collider design.
}
		\label{tab:para}
	\end{center}
\end{table}

\begin{figure}[t]
\begin{center}
\includegraphics[width=10.cm]{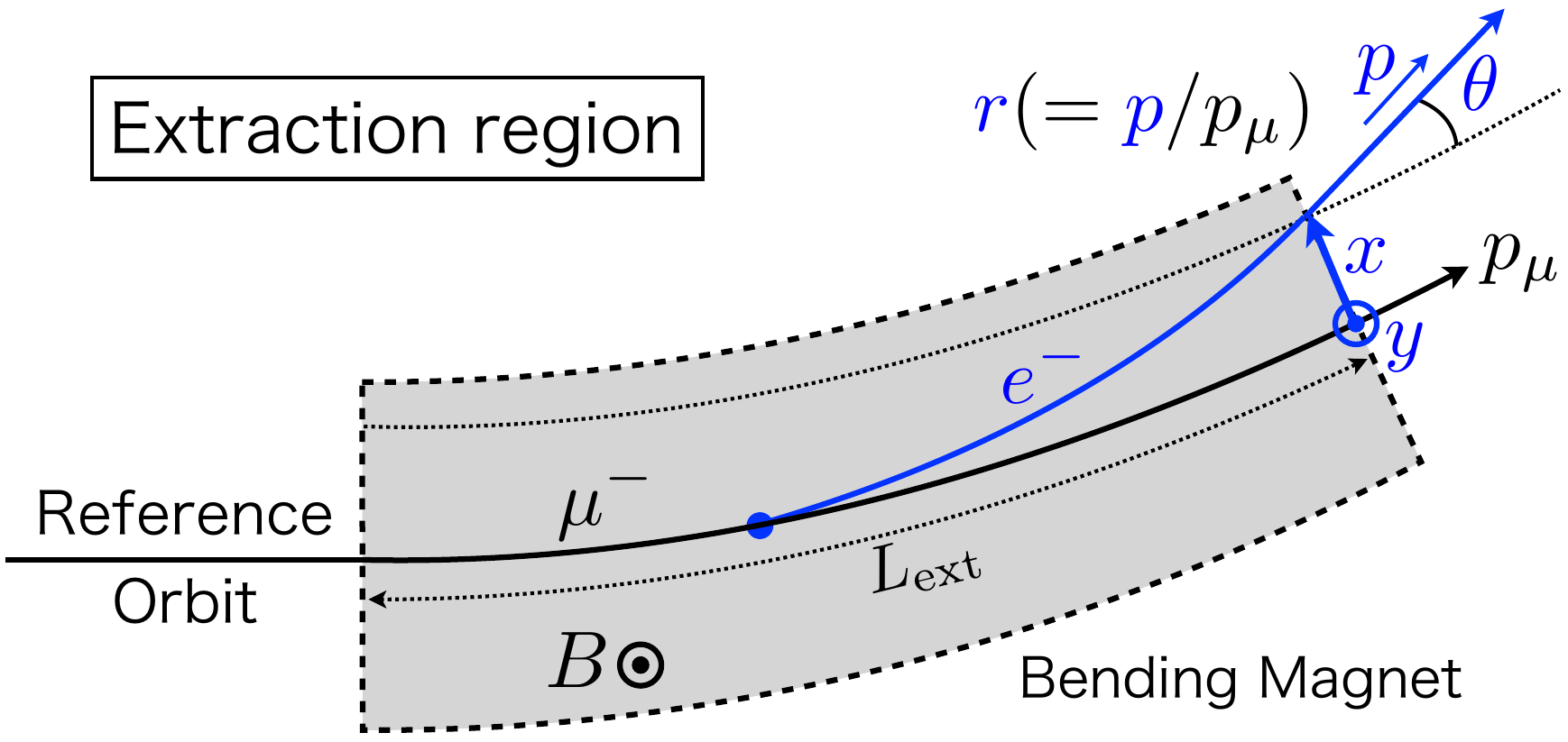}
\end{center}
\vspace{-0.4cm}
\caption{Schematic illustration of the extraction region.
Electrons produced from muon decays are bent by the magnetic field $B$ of the bending magnets (or magnets providing an equivalent magnetic field) installed along the curved section of the collider ring.
The reference orbit denotes the stable muon trajectory in the absence of decay, and the extraction region has length $L_{\rm ext}$.
At the downstream edge of the extraction region, we define the kinematic variables: the electron momentum $\vec{p}$ and the muon momentum in the absence of decay $\vec{p}_\mu$.
The angle $\theta$ is defined, in the horizontal plane, as the angle between $\vec{p}$ and $\vec{p}_\mu$, and the momentum ratio is given by $r=p/p_\mu$ with $p=|\vec{p}|$ and $p_\mu=|\vec{p}_\mu|$.
The transverse deviation from the reference orbit at the downstream edge is denoted by $(x, y)$, where 
$x$ is the horizontal displacement toward the curvature center and $y$ is the vertical displacement.
}
\label{fig:setup}
\end{figure}

Below, we examine the basic principle of extracting electron beams in muon collider experiments.
A typical example of beam extraction in high-energy circular accelerators is the kicker-septum system~\cite{Goddard:2003ni,Goddard:2003dr}, which is used at the LHC to extract 7 TeV proton beams.
At the LHC, the kicker magnet can deflect the proton beam by about 0.27 mrad.
Subsequently, a $\sim$90 m drift section together with quadrupole magnets produces a beam offset of about 3 mm --- sufficient for injection into the septum magnet.
A similar system is also planned for the FCC-hh, aiming to extract proton beams in the 3.3\text{--}50 TeV range~\cite{PhysRevAccelBeams.20.041002}.
In contrast, in muon collider experiments --- our main target --- decay electrons carry lower momenta than their parent muons and therefore experience larger bending in a given magnetic field.
This feature is advantageous, as the bending magnets (or magnets providing an equivalent magnetic field) in the ring may naturally function as a pre-septum magnet without requiring a kicker.
Indeed, as will be explained later, deflections of 0.1\text{--}10 mrad --- significantly larger than in the LHC extraction system --- can typically be achieved, suggesting that such an extraction scheme could be practically feasible and attractive.

Motivated by these considerations, we explore the possibility of utilizing a section of the bending magnets installed along the curved part of the collider ring as an extraction region for the electron/positron beams.
Although the practical extraction of the beam would potentially require dedicated arrangements of magnets and additional apparatus, in what follows we perform an analysis based solely on the geometrical size of the collider ring and the magnetic field strength of the bending magnets, taken as benchmark parameters, with the aim of clarifying the general properties of electron and positron extraction.
Figure~\ref{fig:setup} illustrates schematically the behavior of electrons produced from muon decays within the extraction section.
The reference orbit shown in Fig.~\ref{fig:setup} is defined as the stable muon trajectory ({\it i.e.}, the path a muon would follow in the absence of decay) under the magnetic field.
Once a muon decays in this region, the resulting electron, having lower momentum than the parent muon, is naturally bent inward relative to the reference orbit by the dipole field.
Throughout this paper, we assume unpolarized muon beams.

As shown in Fig.~\ref{fig:setup}, we define the kinematic variables at the downstream edge of the extraction region as follows.
The angle $\theta$ is defined, in the horizontal plane, as the angle between the electron momentum $\vec{p}$ and the tangential direction of the muon reference orbit, characterized by the muon momentum $\vec{p}_\mu$.
The deviation from the reference orbit is denoted by $(x,y)$, where $x$ represents the horizontal direction (toward the curvature center) and $y$ the vertical direction.
Let the magnitudes of the electron and muon momenta be denoted by $p=|\vec{p}|$ and $p_\mu=|\vec{p}_\mu|$, respectively, and define their ratio as
\begin{align}
r=\frac{p}{p_\mu}.
\end{align}
For later convenience, we denote the magnetic field strength in the bending magnet by $B$, and the length of the extraction region by $L_{\rm ext}$.

With this setup, we perform Monte Carlo simulations using PHITS, and the results are presented below.
Figure~\ref{fig:flux} shows the fluxes of particles ($e^-$, $\bar{\nu}_e$, and $\nu_\mu$) produced by $\mu^-$ decays in the extraction region, together with the $\mu^-$ reference orbit.
In this plot, we assume a muon beam energy of 1.5 TeV and a magnetic field strength of 10 T, consistent with the IMCC design, and we adopt $L_{\rm ext}=10$ m as a representative extraction length.
Muons decay uniformly throughout the extraction region, and the resulting electrons are bent toward the positive $x$ direction by the magnetic field.
Neutrinos, on the other hand, are electrically neutral and thus propagate almost tangentially to the reference orbit.
\begin{figure}[t]
\begin{center}
\includegraphics[width=12.cm]{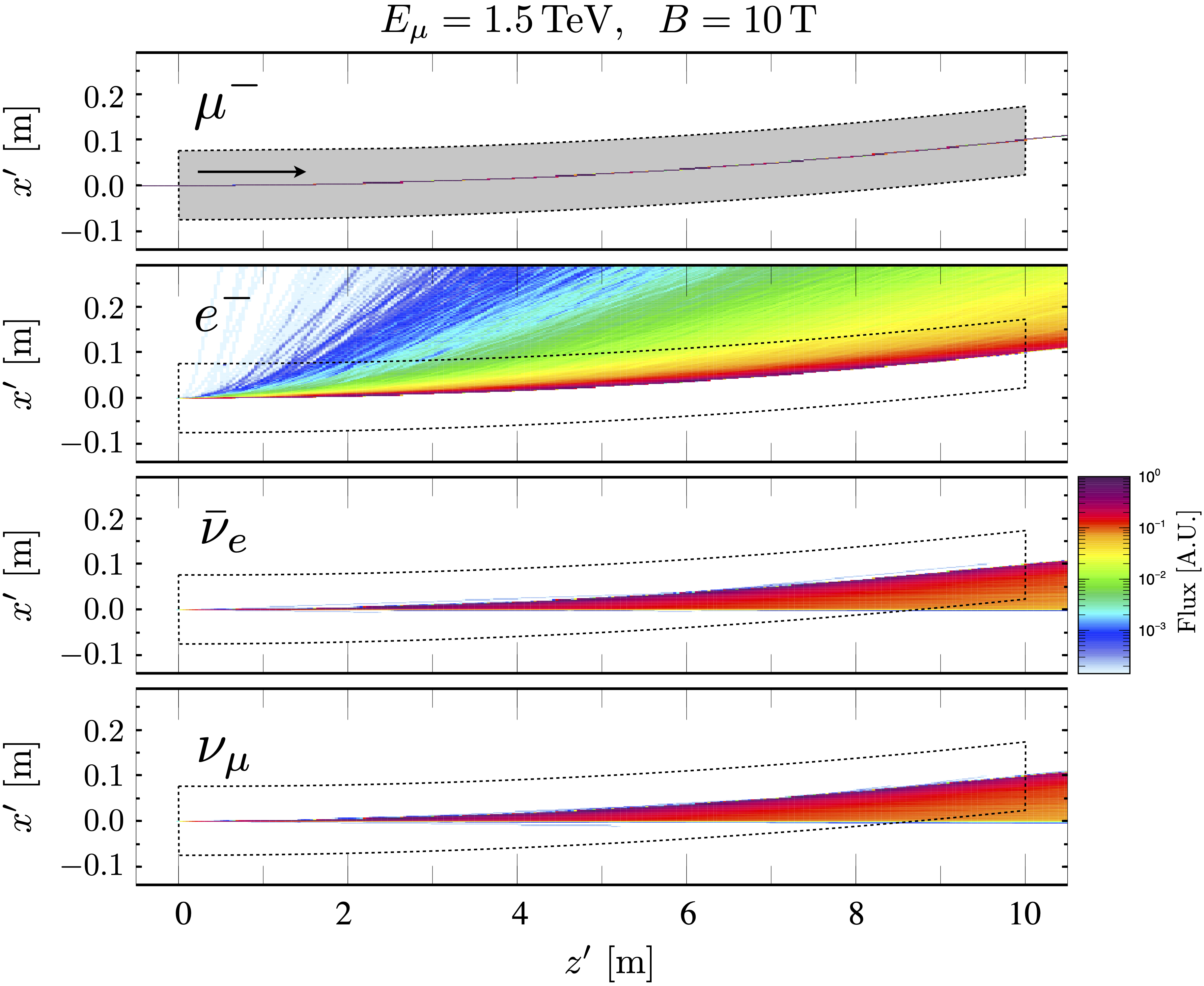}
\end{center}
\caption{Fluxes of particles $e^-$, $\bar{\nu}_e$, and $\nu_\mu$ produced by $\mu^-$ decays in the extraction region, obtained from Monte Carlo simulations using PHITS.
The extraction region enclosed by black dotted lines corresponds to the same slice as in Fig.~\ref{fig:setup}.
For convenience, we newly introduce two axes, $x'$, and $z'$. 
A muon beam energy of 1.5 TeV and a magnetic field strength of 10 T are assumed, and 
$L_{\rm ext}=10$ m is taken as a representative extraction length.
The gray region denotes the effective magnetic field region corresponding to that shown in Fig.~\ref{fig:setup}.
}
\label{fig:flux}
\end{figure}
Moreover, Fig.~\ref{fig:4_1500000_10.0_1} shows the $x$\text{-}$y$ position distributions of the decay products at the downstream edge of the extraction region, obtained from Monte Carlo simulations under the same conditions.
We find that, in contrast to neutrinos, charged electrons are deflected toward the positive $x$ direction.
In particular, electrons with small momenta can be swept away by the magnet, and the resulting width of the $y$ distribution at the downstream edge is suppressed compared with that of neutrinos.
For all decay products, we find that the spread of the $y$-position distribution is much smaller than that of the $x$-position distribution.
Based on this observation, we neglect the dependence on kinematic variables along the $y$-direction throughout this work, which serves as a good approximation.
\begin{figure}[t]
\begin{center}
\includegraphics[width=17.cm]{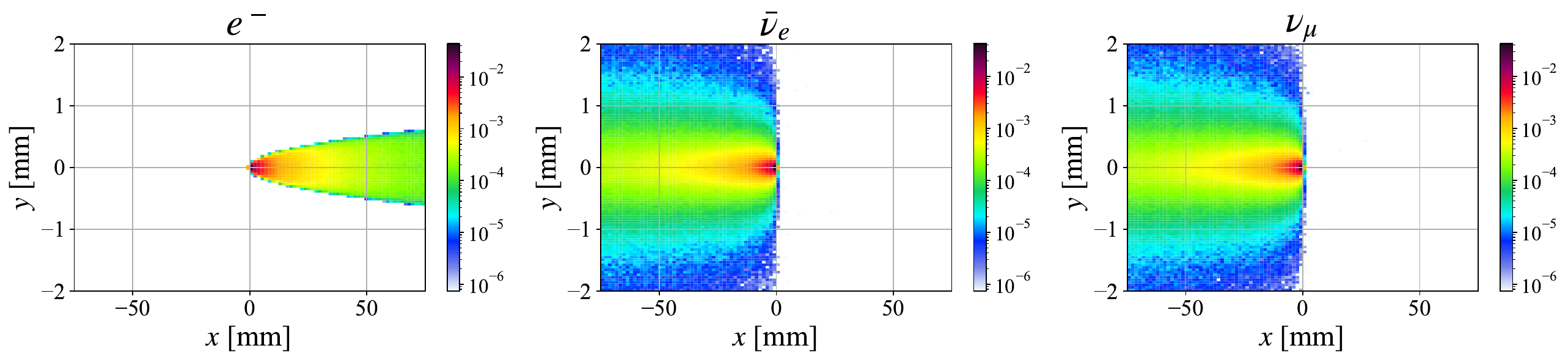}
\end{center}
\vspace{-0.4cm}
\caption{$x$\text{-}$y$ position distributions of decay products ($e^-$, $\bar{\nu}_e$, and $\nu_\mu$) at the downstream edge of the extraction region in Fig.~\ref{fig:setup}, obtained via Monte Carlo simulations. 
We assume the same set of parameters as those in Fig.~\ref{fig:flux}.
}
\label{fig:4_1500000_10.0_1}
\end{figure}

We now estimate the number of electrons that can be extracted from the decay products in this region and transported to beamlines for external use as
\begin{align}
N_e =
N_\mu\frac{L_{\rm ext}}{C_{\rm coll}}\eta,\label{eq:Ne0}
\end{align}
where $N_\mu$ denotes the number of muons delivered to the ring, which can be estimated from the parameters listed in Table~\ref{tab:para} as
\begin{align}
N_\mu= {\rm (Muons/bunch)}\times{\rm (Bunches/train)}\times{\rm (Repetition~rate)}.
\end{align}
The fraction of muons that decay within the extraction region is included in Eq.~\eqref{eq:Ne0} through the factor $L_{\rm ext}/C_{\rm coll}$,
where $C_{\rm coll}$ (3\text{--}15 km) is the collider circumference and $L_{\rm ext}$ is the length of the extraction region.
Moreover, the factor $\eta$ in Eq.~\eqref{eq:Ne0} denotes the transport efficiency of electrons produced in the extraction region, {\it i.e.}, the fraction successfully delivered to external beamlines without loss.
It is defined as
\begin{align}
\eta = \eta_{\rm acc}\times \eta_{\rm trans},\label{eq:eta}
\end{align}
where $\eta_{\rm acc}$ is the probability that an electron produced in the extraction region falls within the acceptance of the transport-beamline phase space, while $\eta_{\rm trans}$ represents the transmission efficiency, namely the probability that an accepted electron reaches the external-beamline experimental area without loss.
For instance, if the collimator gaps are tightened to reduce the beam intensity and suppress beam spread, the transmission efficiency $\eta_{\rm trans}$ decreases.
Therefore, $\eta_{\rm trans}$ should be chosen depending on the intended application; at this stage, we simply impose the general requirement $\eta_{\rm trans} \leq 1$.
As will be discussed later, $L_{\rm ext}$, the length of the extraction region appearing in Eq.~\eqref{eq:Ne0}, is canceled by $\eta_{\rm acc}$, and $N_e$ is independent of $L_{\rm ext}$.

Below, we estimate $\eta_{\rm acc}$ using a simplified approach.
At this stage, the detailed optical design of the transport beamlines has not yet been studied. Therefore, we estimate $\eta_{\rm acc}$ approximately based solely on the phase-space distributions at the downstream end of the extraction region.
As shown in Fig.~\ref{fig:4_1500000_10.0_1}, the divergence in the $y$ direction is much smaller than that in the $x$ direction.
Therefore, throughout our analysis, we neglect the kinematic variables associated with the vertical direction defined in Fig.~\ref{fig:setup}.
Consequently, the relevant kinematic degrees of freedom are reduced to three variables: $(x,\theta,r)$.
Furthermore, for a fixed momentum ratio $r$, the angle $\theta$ is strongly correlated with the horizontal displacement $x$.
As can be obtained analytically, the relation between $x$ and $r$ can be approximated as
\begin{align}
x=6.25\,{\rm mm}
\times
\frac{r}{1-r}
\frac{p_\mu}{\rm 1.5\,TeV/c}
\frac{\rm 10\,T}{B}
\left(\frac{\theta}{\rm 5\,mrad}\right)^2.\label{eq:x_app}
\end{align}
Therefore, we retain only the two variables $(\theta,r)$ and evaluate the acceptance using the decay-electron probability distribution $d^2P_{\mu\to e}/d\theta dr$.

\begin{figure}[t]
\begin{center}
\includegraphics[width=10.cm]{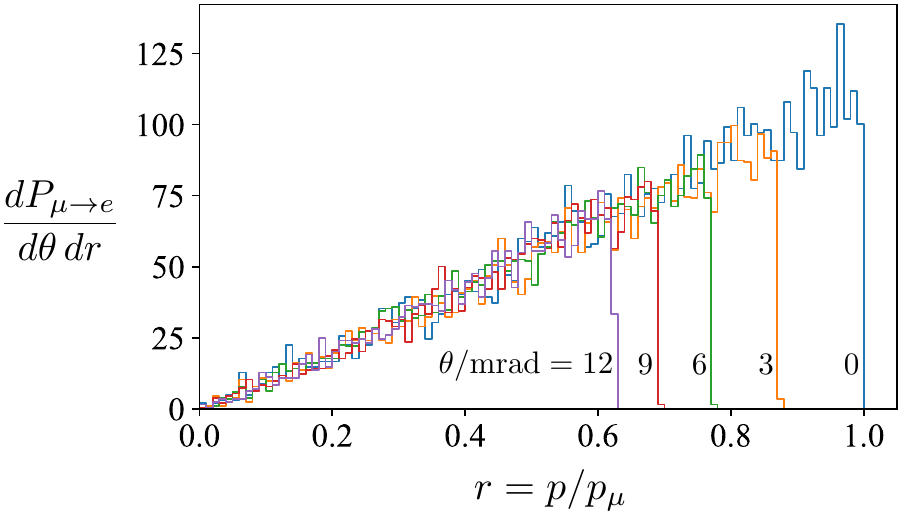}
\end{center}
\caption{Decay-electron probability distributions for the momentum ratio $r=p/p_\mu$ at $\theta = 0,\,3,\,6,\,9,\,12~\mathrm{mrad}$, obtained via Monte Carlo simulations. 
The same parameter settings as in Fig.~\ref{fig:flux} are used.
}
\label{fig:4_1500000_10.0_3}
\end{figure}

By introducing an angular window $\Delta\theta$ and a momentum window $\Delta r$ allowed by the acceptance condition, the acceptance factor is given by
\begin{align}
\eta_{\rm acc} 
= \Delta\theta \, \Delta r \frac{dP_{\mu\to e}}{d\theta \, dr},\label{eq:etaacc}
\end{align}
with
\begin{align}
\frac{dP_{\mu\to e}}{d\theta \, dr}
&\simeq \frac{2r}{\theta_0}\Theta(r_{\rm max}-r),\label{eq:dPdtdr}
\end{align}
where
\begin{align}
\theta_0
= \frac{0.3 \, B{\rm [T]} \, L_{\rm ext}{\rm [m]}}{p_\mu{\rm [GeV/c]}},
~~~
r_{\rm max}
= \frac{\theta_0}{\theta_0+\theta}.\label{eq:theta0}
\end{align}
In particular, because $\theta_0$ contains a factor of $L_{\rm ext}$, Eq.~\eqref{eq:Ne0} becomes independent of $L_{\rm ext}$.
As shown in Fig.~\ref{fig:4_1500000_10.0_3}, these approximated expressions are validated by the PHITS-based Monte Carlo simulation.
The distributions of the momentum ratio $r=p/p_\mu$ are presented for $\theta = 0,\,3,\,6,\,9,\,12~\mathrm{mrad}$, respectively.
In particular, this result indicates that, owing to the bending magnets (or magnets providing an equivalent magnetic field) in muon collider rings, an electron deflection of $0.1~\mathrm{mrad}\text{--}10~\mathrm{mrad}$ --- well above the $\mathcal{O}(0.1~\mathrm{mrad})$ level characteristic of the LHC kicker system --- can be achieved even for high-energy electrons with large momentum fractions $r\simeq 0.6\text{--}1.0$.
This implies that the required drift region for separating the electron beam from the muon beam could be significantly shortened.
As a consequence, neutrino-induced radiation, which is enhanced along the straight sections of the ring, can be significantly reduced.

From Eq.~\eqref{eq:x_app}, the acceptance window in the horizontal displacement, $\Delta x$, can be approximately\footnote{The contributions proportional to $\Delta r/r$ are much smaller than those proportional to $\Delta\theta/\theta$, and can therefore be neglected.} expressed in terms of the allowed angular window $\Delta\theta$ as
\begin{align}
\Delta x \simeq \frac{2x\Delta\theta}{\theta}.\label{eq:Deltax}
\end{align}
For the missing-momentum search considered in Sec.~\ref{sec:DM}, we adopt a momentum acceptance of $\Delta p/p=1\,\%$ as a benchmark.
Here, $p$ denotes the central momentum and $\Delta p$ the full width of the accepted interval.
For comparison, the CERN SPS North Area H4 beamline specifies a maximum momentum acceptance of $\pm1.4\,\%$, corresponding to a full width of $2.8\,\%$~\cite{CERN_H4_beamline}.
By combining Eqs.~\eqref{eq:x_app}\text{--}\eqref{eq:theta0} with Eq.~\eqref{eq:Ne0}, we obtain the following expression for the number of extracted electrons:

\begin{align}
N_e
=
\hat{N}_e
~\frac{\Delta\theta}{\rm 1\,mrad}
~\frac{\Delta p/p}{\rm 1\,\%}
~\frac{10\,{\rm T}}{B}
~r^2
~\eta_{\rm trans}
~\Theta(r_{\rm max}-r),\label{eq:Ne}
\end{align}
where $\Delta p/p=\Delta r/r$.
The reduction in electron yield due to the narrow momentum selection is already included through $\eta_{\mathrm{acc}}$, as reflected in the explicit dependence on $\Delta p/p$ in Eq.~\eqref{eq:Ne}. 
Section~\ref{sec:DM} adopts this same momentum acceptance, while additional transport losses are parametrized by $\eta_{\mathrm{trans}}$.
We note that a typical parameter set
($\Delta\theta = 1\,\mathrm{mrad}$, $\Delta p/p = 1\,\%$, $B = 10\,\mathrm{T}$) is considered here.
In this context, $\hat{N}_e$ denotes the maximum number of electrons obtained for this parameter set ({\it i.e.}, the number of electrons before including the transmission efficiency factor $\eta_{\rm trans}$), and is given by
\begin{align}
\hat{N}_{e} = 
10^{-6}N_\mu
\times
~\frac{\rm 10\,km}{C_{\rm coll}}
~\frac{p_\mu}{\rm 1.5\,TeV/c}.\label{eq:hatNe}
\end{align}
For convenience, the values of $\hat{N}_e$ for each experimental proposal are summarized in Table~\ref{tab:para}.
Notably, $\hat{N}_e$ is independent of $L_{\rm ext}$.
Throughout this paper, one year of operation is taken to be $1.2\times 10^7$ sec. 
These results indicate that $10^{14}$\text{--}$10^{15}$ electrons can be extracted from the extraction region during one year of operation.
In the following sections, we assume $10^{14}$\text{--}$10^{15}$ electrons, depending on the scenario, and consider searches for new light particles.

We next illustrate the time structure of the generated electron beam and its electron yield.
Since the electron beam is potentially generated each time the muon beam passes through the extraction region, the repetition rate of the electron/positron beam is given by
\begin{align}
\text{(Repetition rate of $e^\pm$}) = \frac{c}{C_{\rm coll}}\times\text{(Bunches/train)}.\label{eq:rep_def}
\end{align}
For each experimental proposal, the corresponding values are summarized in Table~\ref{tab:para}, indicating that high-frequency electron/positron beams, ranging from a few tens of kHz to a few MHz, can be achieved in muon collider experiments.
The number of electrons generated per muon bunch crossing of the extraction region is given by
\begin{align}
\text{($e^\pm$/bunch)} 
&= 
\text{(Muons/bunch)}
\times
\frac{\delta t}{\gamma\tau_\mu}e^{-\frac{t}{\gamma\tau_\mu}}
~ \eta,\\
&=
\hat{n}_{e^\pm/{\rm bunch}}
\times
e^{-\frac{t}{\gamma\tau_\mu}}
\frac{\Delta\theta}{\rm 1\,mrad}
~\frac{\Delta p/p}{1\,\%}
~\frac{\rm 10\,T}{B}
~r^2
~\eta_{\rm trans}
~\Theta(r_{\rm max}-r),\label{eq:ebunch}
\end{align}
where $\delta t = L_{\rm ext}/c$ denotes the time during which the muon bunch traverses the extraction region, $t$ is the time elapsed between the injection of the muon beam into the collider ring and its crossing of the extraction region, $\gamma\, (= E_\mu/m_\mu)$ is the Lorentz factor of the muon with mass $m_\mu$, and $\tau_\mu$ is the muon lifetime.
Note that the number of electrons produced from a single muon bunch depends on the time $t$, because the number of muons circulating in the collider ring decreases with time due to muon decay.
The number $\hat{n}_{e^\pm/{\rm bunch}}$ is expressed as
\begin{align}
\hat{n}_{e^\pm/{\rm bunch}}
&=
\text{(Muons/bunch)}\times
\frac{10^{-6}}{0.15}
~\frac{\rm 1\,m}{c\tau_\mu}
~\frac{m_\mu}{\rm GeV}
\\
&\simeq
\text{(Muons/bunch)}\times
1.1\times 10^{-9},\label{eq:hatnperbunch}
\end{align}
The quantity $\hat{n}_{e^\pm/{\rm bunch}}$ represents, for a typical parameter set ($\Delta\theta = 1\,\mathrm{mrad}$, $\Delta p/p = 1\,\%$, $B = 10\,\mathrm{T}$), the maximum number of electrons produced per muon bunch crossing of the extraction region, {\it i.e.}, before including the transmission efficiency factor $\eta_{\rm trans}$.
Note that, when including additional factors such as the transmission efficiency $\eta_{\rm trans}$, the value of $\text{($e^\pm$/bunch)}$ can be reduced, and $\text{($e^\pm$/bunch)}\sim\mathcal{O}(1\text{--}10)$ is also possible.
The estimated values of $\hat{n}_{e^\pm/{\rm bunch}}$ for each experiment are summarized in Table~\ref{tab:para}.
From Table~\ref{tab:para}, we find that, before accounting for the efficiency factor~\eqref{eq:eta}, the expected number of extractable electrons over one year of operation~\eqref{eq:hatNe} is largely independent of the experimental configuration.
In contrast, the IMCC provides a much larger number of electrons per muon bunch~\eqref{eq:hatnperbunch}, whereas $\mu$TRISTAN offers a significantly higher repetition rate.
These differences indicate that the optimal applications of the extracted electron/positron beams depend strongly on the specific muon collider design.
As will be discussed later in Sec.~\ref{sec:miss}, in the context of searches for new light particles, the $\mu$TRISTAN setup is more suitable for missing energy/momentum searches, whereas the IMCC setup is more suitable for visible decay searches.

In the analysis above, we assumed a uniform dipole magnetic field over the extraction region.
In a realistic accelerator, however, the curved sections will be populated with a strong-focusing alternating-gradient lattice that includes quadrupole and sextupole magnets in addition to the main dipoles.
Within a quadrupole magnet, the magnetic field depends linearly on the transverse position, so that a muon decaying in such a region would produce an electron experiencing a position-dependent deflection rather than the uniform bending assumed in Eq.~\eqref{eq:x_app}.
This would smear the idealized $(x,\theta)$ phase-space correlation derived above and distort the angular acceptance.
Sextupole fields, being higher-order corrections, have a negligible effect near the beam axis.
We note, however, that the placement of the extraction region is a design parameter: it is natural and practical to locate the extraction region within a section dominated by dipole magnets, where the uniform-field approximation adopted in this work remains valid.
Importantly, because the number of extractable electrons $N_e$ is independent of the extraction length $L_{\rm ext}$ (see Eqs.~\eqref{eq:Ne}\text{--}\eqref{eq:hatNe}), restricting the extraction region to a dipole-dominated section does not alter the order-of-magnitude estimates presented above.
Furthermore, quadrupole magnets do not necessarily degrade the extraction performance; with an appropriate lattice design, they can enhance it.
At the LHC, for example, the total horizontal deflection of the beam dumping system is quoted as a combined ``MKD~+~Q4'' contribution~\cite{Evans:2008zzb}, and in the LHC injection system, a defocusing quadrupole placed between the kicker and septum magnets provides about one quarter of the required total deflection through kick enhancement~\cite{Evans:2008zzb}.
A detailed optimization of the phase-space acceptance based on a realistic lattice design is left for future work.

\section{New light particle search}
\label{sec:DM}

\begin{figure}[t]
\begin{center}
\includegraphics[width=6.cm]{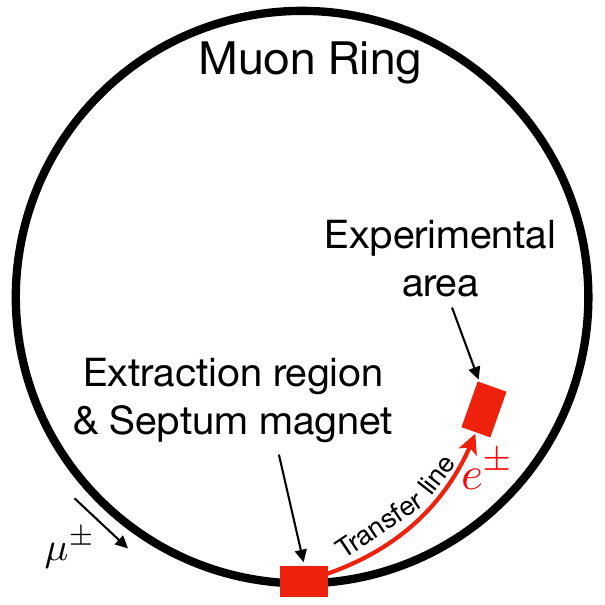}
\end{center}
\vspace{-0.4cm}
\caption{Schematic of the connection between the extraction region and the experimental area for new light particle searches. 
The electron/positron beams extracted from the region of Fig.~\ref{fig:setup} are transferred to the experimental area, for example using septum magnets.}
\label{fig:ext}
\end{figure}

To illustrate the importance of carefully considering positron/electron beam extraction in muon collider experiments, we examine the potential for new light particle searches.
As discussed above, the properties of the extracted beams depend on the specific collider design; we therefore consider two search strategies in what follows.
In Sec.~\ref{sec:miss}, we study a missing energy/momentum search at the $\mu$TRISTAN beam, focusing on sub-GeV dark matter interacting via a dark photon.
We employ an LDMX-like setup with continuous $\mu$TRISTAN beams and evaluate its sensitivity to sub-GeV dark matter.
In Sec.~\ref{sec:visible}, we explore a visible-decay search at the IMCC beam, considering axion-like particles (ALPs) and scalars coupled to photons.
Unless stated otherwise, we take $r=0.8$ --- corresponding to the maximum deflection angle $\theta = 5~\mathrm{mrad}$, about twenty times larger than that of LHC beam extraction --- as motivated by Fig.~\ref{fig:4_1500000_10.0_3}.
For concreteness, as illustrated in Fig.~\ref{fig:ext}, we assume that the extracted positron/electron beams are guided by downstream magnets --- such as a high-field septum magnet --- to an experimental setup for new light particle searches.

In muon collider experiments, decay products of muon beams strike beamline components throughout the ring, producing secondary particles --- notably highly penetrating neutrons and muons --- that can constitute a potential background for the searches discussed below.
These secondary particles can reach the experimental area either by propagating along the transfer line or by penetrating through tunnel walls and rock.
For the latter path, Monte Carlo studies~\cite{Sakaki:2022udd} show that the neutron flux is attenuated by approximately one order of magnitude per meter of concrete shielding, so that several meters of intervening rock provide sufficient attenuation.
For the former path, bending the transfer line with dipole magnets is an effective mitigation strategy.
Because the extracted electrons and positrons have lower momenta than the parent muons, they can be bent more strongly, allowing the experimental area to be well separated from the main ring.
Bending the transfer line eliminates the line of sight to the ring, so that forward-going neutrons are absorbed in the tunnel walls; additional shielding structures placed along the bent sections can further reduce the residual flux.
Secondary muons produced along the ring are strongly forward-peaked in the tangential direction~\cite{Sakaki:2022udd} and have low energy and penetrating power in the transverse direction; placing the experimental area outside the horizontal plane of the collider ring further suppresses the muon flux reaching the detectors.
Based on these considerations, we expect that this background can be reduced to a negligible level by appropriate transfer-line and shielding design, and we do not include it in the analyses below.

\subsection{Missing energy/momentum search in the $\mu$TRISTAN setup}
\label{sec:miss}
The $\mu$TRISTAN positron/electron beam exhibits a continuous high-energy structure. 
When the transfer efficiency $\eta_{\rm trans}$ (see Eq.~\eqref{eq:eta}) is taken into account, only $\mathcal{O}(1\text{--}10)$ electrons are expected per bunch, despite the high repetition rate.
Such a beam configuration is particularly suitable for searches for missing energy/momentum signatures of new particles.
In what follows, focusing on sub-GeV dark matter, we investigate missing energy/momentum searches using the $\mu$TRISTAN positron/electron beam.

\subsubsection{Detectable DM models}\label{sec:model}
As a benchmark model, we consider a class of DM scenarios in which the DM field is charged under a new gauge symmetry U(1)$_X$ at a high energy scale.
In particular, we study the corresponding low-energy theory in which U(1)$_X$ is spontaneously broken and the hidden photon $X$ acquires a mass $m_{X}$.
The relevant terms of the Lagrangian are then given by:
\begin{align}
    \mathcal{L}\supset -\frac{1}{4}F_{\mu\nu}F^{\mu\nu} -\frac{1}{4} X_{\mu\nu} {X}^{\mu\nu} +\frac{1}{2} m^2_{X} X_{\mu} {X}^{\mu}
    -\frac{\epsilon}{2}X_{\mu\nu}F^{\mu\nu}-g_D X_{\mu} J^{\mu}_{\chi}-eA_{\mu} J^{\mu}_{\rm EM},\label{eq:Lag_DM}
\end{align}
where $F_{\mu\nu}$ ($X_{\mu\nu}$) denotes the photon (dark photon) field strength, $J^{\mu}_{\rm EM}$ is the SM electromagnetic current, and $g_D$ is the couplings.
Moreover, we can include DM candidates $\chi$ charged under U(1)$_X$, with $J^{\mu}_{\chi}$ denoting the corresponding DM current, {\it e.g.},
\begin{align}
    J^{\mu}_{\chi}=\begin{cases}
    i\overline{\chi}_2 \gamma^{\mu}\chi_1 +{\rm H.c.}~~~~~~~\text{Pseudo-Dirac DM}
    \\
    i \chi^{\ast}\partial^{\mu}\chi+{\rm H.c.}~~~~~~~~\,\text{Scalar elastic DM}
    \\
    \chi_1\partial^{\mu}\chi_2-\chi_2\partial^{\mu}\chi_1~~~~\text{Scalar inelastic DM}
    \\
    \frac{1}{2}\overline{\chi} \gamma^{\mu}\gamma_5 \chi~~~~~~~~~~~~~~~\text{Majorana DM}
    \end{cases}\label{eq:DMcu}
\end{align}
The parameter $\epsilon$ represents the kinetic mixing between the SM photon and the dark photon $X$.
By redefining $A_{\mu}\to A_{\mu}-\epsilon X_{\mu}$, the gauge kinetic terms are canonically normalized to first order in $\epsilon$, and the interaction terms are then given by:
\begin{align}
    \mathcal{L}_{\rm int}= -g_D X_{\mu} J^{\mu}_{\chi}+\epsilon e X_{\mu} J^{\mu}_{\rm EM} -e A_{\mu} J^{\mu}_{\rm EM}.
\end{align}
In particular, as will be explained below, the third term above gives rise to a missing-momentum signature in the LDMX-like setup.
For the on-shell dark photon scenario considered here, within the parameter range of interest, the missing-momentum signature is approximately insensitive to the detailed structure of the DM current in Eq.~\eqref{eq:DMcu}.

\subsubsection{Experimental setup}

\begin{figure}[t]
\begin{center}
\includegraphics[width=11.cm]{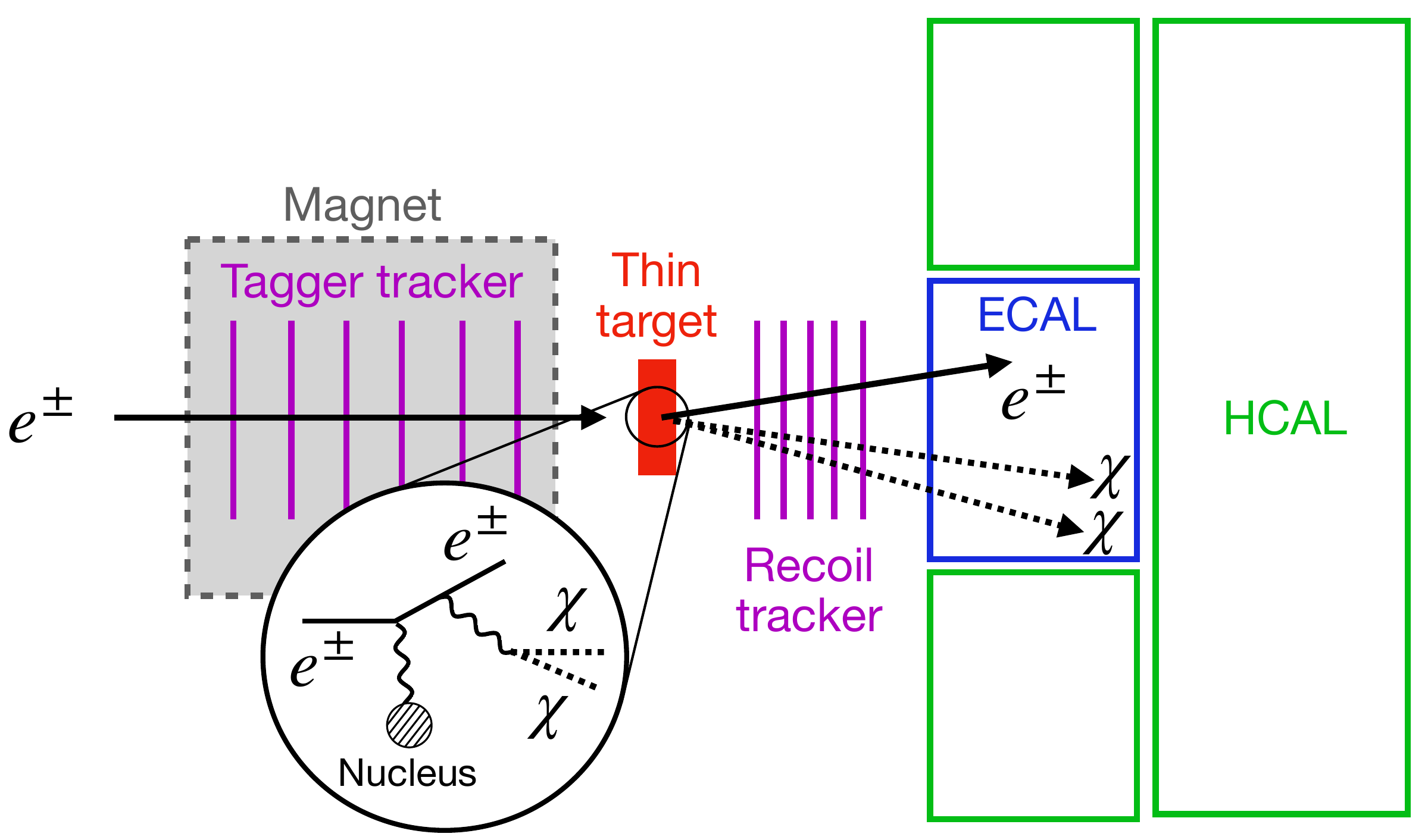}
\end{center}
\vspace{-0.4cm}
\caption{The LDMX experimental~\cite{Akesson:2022vza} setup mainly consists of a tagger tracker in a magnetic field region, a thin tungsten target, a recoil tracker, and an electromagnetic calorimeter (ECAL) and a hadronic calorimeter (HCAL).
This schematic is based on the LDMX detector layout shown in Ref.~\cite{Akesson:2022vza}.
The extracted positron/electron beams from the muon collider pass through the tagger tracker, where their momenta are measured before entering the thin target.
DM particles may be produced at the thin target, {\it e.g.}, via bremsstrahlung.
The momentum of the final-state positron/electron is measured by the recoil tracker, allowing the reconstruction of the missing momentum associated with the DM particles.
The ECAL and HCAL veto photons and hadrons that could mimic missing energy signatures.
}
\label{fig:LDMX}
\end{figure}

In the following analysis, we assume that the extracted electron beam carries $r=80\,\%$ of the muon beam energy.
For the $\mu$TRISTAN setup, this corresponds to an electron beam energy of $E_{\rm beam}=800~\mathrm{GeV}$ (see Table~\ref{tab:para}).
For ease of comparison with LDMX Phase~1, assuming $4\times10^{14}$ electrons on target at a 4~GeV electron beam energy, we take $N_e = 4\times10^{14}$ electrons accumulated over an operational period of $\mathcal{O}(1\text{--}10)$ years in our analysis.
To illustrate the potential applications of continuous positron/electron beams, we consider a DM search.
In contrast to bunched beams, continuous beams can be tagged even in fixed-target experiments, making missing momentum/energy searches one of the most promising approaches.
We adopt an experimental setup similar to that of the LDMX experiment in Ref.~\cite{Akesson:2022vza}, as illustrated in Fig.~\ref{fig:LDMX}.
For instance, as shown in Fig.~\ref{fig:ext}, we assume that the continuous positron/electron beams are extracted, and then sent to an LDMX-like setup by subsequent magnets.
A tagger tracker placed within a magnetic field is assumed to be located in front of a thin tungsten target to precisely characterize the initial positron/electron beam and its momentum.
Following the tagger tracker, a thin tungsten target with a thickness of $l_{\rm target} = 0.1\, X_0$ (with $X_0=0.3504$ cm~\cite{ParticleDataGroup:2024cfk}) is placed.
Behind the thin target, a second tracker is assumed to reconstruct the momentum of the final-state positron/electron and identify the missing momentum associated with the DM particles.
To suppress background events such as high-energy photon production, muon pair production, and neutral hadron production, an electromagnetic calorimeter (ECAL) and a hadronic calorimeter (HCAL) are placed downstream of the second tracker.
The tagging tracker must be adapted to the substantially higher incident momentum considered here.
For $n$ equally spaced position measurements in a uniform magnetic field, the contribution of finite spatial resolution to the relative momentum resolution is approximately~\cite{Drasal:2018zij}
\begin{align}
\left(\frac{\sigma_p}{p}\right)_{\mathrm{meas}}
\simeq
\frac{\sigma_x p}{0.3\,B_{\mathrm{tag}}L^2}
\sqrt{
\frac{720(n-1)^3}
{(n-2)n(n+1)(n+2)}
}.
\end{align}
Here, $B_{\mathrm{tag}}$ is the magnetic field perpendicular to the bending plane, $\sigma_x$ is the single-hit position resolution in the bending direction, and $L$ is the distance between the first and last measurement planes within the assumed uniform field. 
The units are $p$ in GeV/$c$, $B_{\mathrm{tag}}$ in tesla, and $L$ and $\sigma_x$ in metres.
As an illustrative benchmark, we retain the nominal field strength and tracking-layer parameters of the LDMX design, $B_{\mathrm{tag}}=1.5$ T, $n=7$, and $\sigma_x=6~\mu\mathrm m$~\cite{LDMX:2018cma}, while extending the measurement span to $L=3.2$ m.
At $p=800$ GeV/$c$, the above expression gives
$
\left({\sigma_p}/{p}\right)_{\mathrm{meas}}
\simeq0.82\%.
$
For a representative material budget of $0.7\%\,X_0$ per layer, motivated by the HPS tracker discussed in Ref.~\cite{LDMX:2018cma}, the accumulated multiple-scattering angle is approximately $3.3~\mu\mathrm{rad}$.
Comparing this with the magnetic bending angle of approximately $1.8~\mathrm{mrad}$ gives a rough estimate of $0.2\%$ for the corresponding relative momentum-resolution contribution.
Its precise value depends on the layer geometry and track-fitting procedure.
These estimates illustrate the magnetic measurement span required for percent-level tracking resolution under the stated assumptions. 
They do not establish the complete detector performance, which requires a dedicated assessment of the magnetic field map, alignment, detector material, bremsstrahlung, and synchrotron-radiation effects.
This benchmark concerns the upstream tagging tracker; the recoil-tracker performance requires a separate assessment.

Now, we consider the interaction of a continuous positron/electron beam with target particles in a thin target, producing a dark photon, $X$, which subsequently decays into DM particles.
As a feasibility study, this paper focuses on the on-shell production of $X$ via bremsstrahlung, as illustrated in Fig.~\ref{fig:LDMX}.
The number of signal events for bremsstrahlung $e^{\pm}{\rm N}\to e^{\pm} X{\rm N}$ with a target nucleus N is then given by
\begin{align}
    N_{\rm signal}=  N_{e}\cdot n_{\rm N} \cdot l_{\rm target} \int dE_{\rm recoil} \frac{d \sigma \left(e^{\pm}{\rm N}\to e^{\pm} X{\rm N}\right)}{d E_{\rm recoil}}\cdot {\Theta}\left(E_{\rm cut}-E_{\rm recoil}\right)\cdot {\rm Br}\left(X\to \chi\bar{\chi}\right),\label{eq:mis_sig}
\end{align}
where $N_{e}=4\times 10^{14}$ is the number of incoming beam electrons, $n_{\rm N}$ is the number density of nucleus N, $l_{\rm target}$ is the length of the thin target, $E_{\rm recoil}$ is the recoil energy of the final state positron/electron, and $\sigma$ is the production cross section of $X$.
Since large missing momentum/energy characterizes the DM signature, we impose an energy cut, $E_{\rm cut}$, on the final-state positron/electron.
In the parameter range of interest, the dark photon $X$ decays approximately exclusively into DM-sector particles, {\it i.e.}, ${\rm Br}(X \to \chi\bar{\chi}) = 1$.
In the following analysis, we consider electrons and positrons selected within a narrow momentum interval centered at $p=800$ GeV/$c$, with a full width of $\Delta p/p=1\,\%$.
This is the same momentum acceptance used in Sec.~\ref{sec:cont}, where the associated reduction in electron yield is included in Eqs.~\eqref{eq:etaacc} and \eqref{eq:Ne}.
Momentum selection can be implemented using bending magnets, focusing magnets, and collimators. 
Particles with different momenta follow different trajectories in a magnetic field, allowing a suitably placed collimator to select a narrow momentum interval.
The H4 beamline provides an established example of this principle: electrons and positrons produced with a broad momentum distribution are selected and transported within a narrow interval around a chosen central momentum~\cite{CERN_H4manual}.
However, the H4 precedent does not establish the achievable performance at 800 GeV.
Determining whether the assumed momentum interval and delivered flux can be realized simultaneously requires a dedicated beam-optics and transport study accounting for the extracted beam’s phase-space distribution, magnet apertures, and synchrotron-radiation losses and fluctuations.
Such a study lies beyond the scope of the present work.
The sensitivities presented below are therefore benchmark projections conditional on the stated beam properties and electron exposure.
By installing magnets and a synchrotron radiation detector in front of the tagger tracker --- similar to those used in the NA64 experiment~\cite{Depero:2017mrr,NA64:2017vtt,Banerjee:2019pds} at the H4 beam line in the North Area (NA) of the CERN SPS --- the identification of positrons and electrons against beam contamination is also expected to be improved.

Prior to presenting numerical results, we examine potential sources of background (BG) events.
The BG events in the missing momentum search are primarily limited to beam-induced processes.
The main reducible BG sources in the missing momentum search include: (i) misidentified initial beam particles, and (ii) photon-nuclear reactions, as well as photon and muon production processes.
BG (i) typically arises from impure positron/electron beams containing contaminations such as charged hadrons or low-energy positrons/electrons.
These impurities affect the preparation of the initial beam momentum and can mimic missing momentum signals.
BG (ii) arises from SM processes occurring at the thin target when struck by the positron/electron beam, which can also mimic missing momentum signatures.
For instance, high-energy photons produced at the thin target may go undetected by the downstream tracker and be misidentified as missing momentum.
The above BG sources can be significantly reduced, though the extent depends strongly on the experimental setup.
BG (i) can be suppressed, for example, by precisely measuring synchrotron radiation induced by the beam particles, combined with accurate momentum tagging using the tagger tracker and well-prepared monochromatized beams.
BG (ii) can be mitigated by placing ECAL and HCAL detectors behind the thin target in addition to the recoil tracker.
Given that these BG reduction strategies are highly dependent on the experimental setup, we assume throughout this paper that these background events are negligible.
For further details on BG reduction strategies, see, for example, Ref.~\cite{LDMX:2019gvz}.

The irreducible BG events correspond to SM processes involving a single positron or electron track accompanied by neutrinos.
Irreducible BG sources include neutrino trident-like processes, such as $e^\pm {\rm N} \to e^\pm \nu_e \bar{\nu}_e {\rm N}$, and a combination of M{\o}ller scattering ($e^\pm e^- \to e^\pm e^-$) followed by a charged-current quasi-elastic (CCQE) reaction ($e^- p \to n \nu_e$).
As evaluated in Appendix~\ref{app:Irre}, for the beam flux adopted in the following analysis, these rates are negligible.

We now briefly discuss the pile-up effects of the electron/positron beams on the number of signal events.
Since $\hat{n}_{e^\pm/{\rm bunch}}=24$ as shown in Table~\ref{tab:para}, reducing the transmission efficiency ({\it i.e.}, tightening the collimator gaps) allows for a beam configuration in which each extracted bunch contains $\mathcal{O}(1\text{--}10)$ electrons.
Even for 2 electrons per bunch, consistent with the LDMX design, $N_e=4\times 10^{14}$ can be achieved for 4--8 years.
In missing-momentum/energy searches, the presence of multiple electrons makes the reconstruction of events with missing momentum or energy challenging.
Assuming that the number of electrons $n$ per bunch follows a Poisson
distribution,
$P(n;\bar{n})=\bar{n}^n e^{-\bar{n}}/n!$,
where $\bar{n}$ is the mean number of electrons per bunch, the probability
of having exactly one electron in a bunch is $P(1;\bar{n})$.
For a conservative estimate, we assume that only bunches containing
exactly one electron can be used for the signal search.
For $N_b$ bunches incident on the target, the expected total number of
incident electrons is $N_e=N_b\,\bar{n}$, whereas the expected number of
accepted electrons is $N_{\rm acc}=N_b\,P(1;\bar{n})$.
Therefore, the signal efficiency associated with this single-electron requirement is $\eta_{\rm pu}\equiv N_{\rm acc}/N_e = e^{-\bar{n}} \simeq 0.135$ for $\bar{n}=2$.
For a mass-independent pile-up efficiency \(\eta_{\rm pu}\), the background-free sensitivity in \(y\) scales as \(y_{\rm lim}(\eta_{\rm pu})=y_{\rm lim}(1)/\eta_{\rm pu}\). The single-electron requirement at fixed \(\bar n=2\) would therefore shift the sensitivity curve upward by a factor of approximately \(7.4\). Figure 8 retains the ideal projection with \(\eta_{\rm pu}=1\).
As discussed below, techniques developed for LDMX may help mitigate pile-up effects and improve the efficiency beyond this conservative estimate. 
Motivated by these possibilities, we do not apply this conservative efficiency factor to the sensitivity projections in Fig.~8, which are presented as benchmarks assuming negligible residual losses from pile-up.

However, in the LDMX experiment, several strategies to mitigate pile-up effects have been studied.
For example, Ref.~\cite{LDMX:2025bog} presents various pile-up mitigation techniques, including beam spreading to reduce event occupancy (Table~3.1 in Sec.~3.3.3), a multi-electron trigger that optimizes the missing-energy cut depending on the number of incoming electrons (Table~4.9 in Sec.~4.3.3), and a machine-learning-based ECal veto for multi-electron events (Sec.~4.3.3).
The total signal efficiency is shown to reach 74\% even for an average of two electrons per bunch (Table~4.9 in Sec.~4.3.3).
Although our experimental setup differs from that of LDMX in terms of beam energy and detector design, preventing a direct comparison, these methods may still help mitigate pile-up effects in our setup.
In light of these considerations, we defer a detailed analysis of the signal efficiency in the presence of pile-up effects to future work and therefore omit it from the sensitivity estimates.

\subsubsection{Numerical results}
We now present the results of Monte Carlo simulations performed using the PHITS code.
To perform numerical simulations for new physics searches in our experimental setup, we have implemented the exact calculation\footnote{The calculation does not rely on approximations such as the Weizsacker\text{-}Williams (WW) approximation.
We have verified that our numerical results agree with those obtained using the complete calculation presented in Refs.~\cite{Liu:2016mqv,Liu:2017htz}.
} of the dark photon bremsstrahlung process in PHITS.
The details of the PHITS code implementation, including the exact bremsstrahlung processes of the new particles, will be presented in a separate publication.

\begin{figure}[t]
\begin{center}
\includegraphics[width=15.cm]{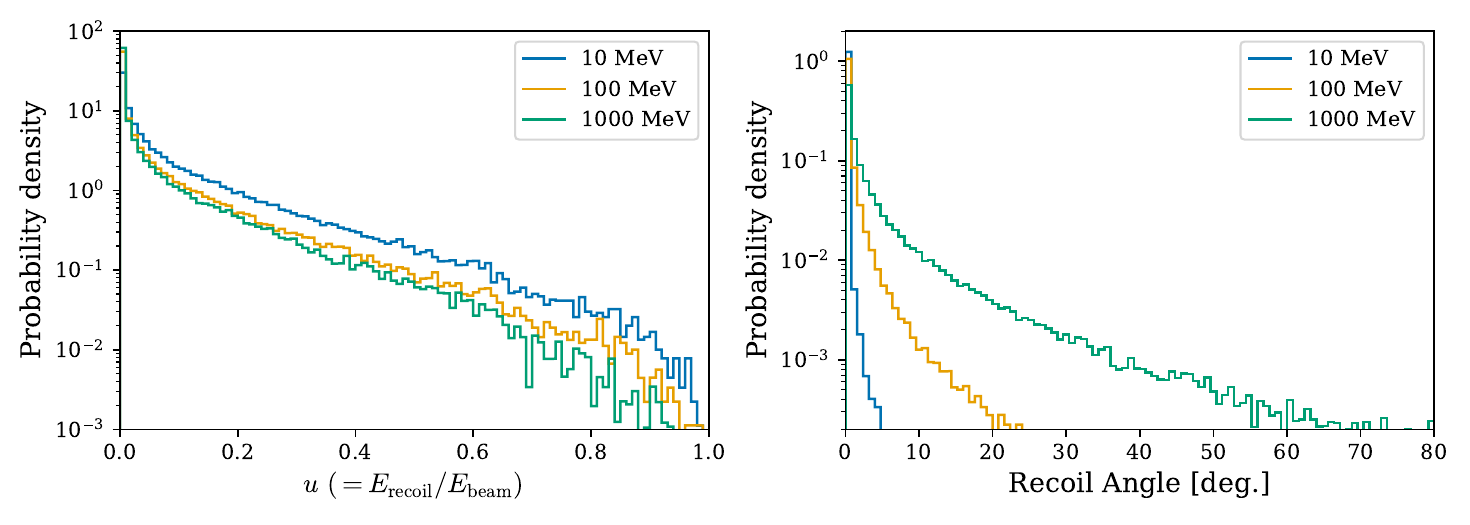}
\end{center}
\vspace{-0.4cm}
\caption{Probability density functions for the normalized recoil energy $u = E_{\rm recoil}/E_{\rm beam}$ (left) and the recoil angle (right).
The differently colored lines correspond to dark photon masses
$m_{A'} = 10~{\rm MeV}, 100~{\rm MeV}$, and $1000~{\rm MeV}$.
}
\label{fig:dis}
\end{figure}

For the signal process $e^{-}{\rm N}\to e^{-} X{\rm N}$, Fig.~\ref{fig:dis} shows the probability density functions of the recoil energy of the final-state electron (left panel), expressed in terms of the normalized variable $u=E_{\rm recoil}/E_{\rm beam}$, and the recoil angle (right panel).
The differently colored lines correspond to dark photon masses $m_{A'}=$ 10 MeV, 100 MeV, and 1000 MeV.
From the left panel of Fig.~\ref{fig:dis}, we find that for small values of the dark photon mass $m_{A'}$, the recoil-energy distributions extend to higher-energy regions.
Since the potentially dominant BG process $e^{-}{\rm N}\to e^{-} \gamma {\rm N}$ corresponds to the limit $m_{A'}=0$, this BG can be efficiently suppressed by imposing an upper cut on the recoil energy, $E_{\rm recoil}\leq  E_{\rm cut}$.
Moreover, the dependence on the beam energy $E_{\rm beam}$ of the probability density functions, expressed as functions of $u = E_{\rm recoil}/E_{\rm beam}$, is mild for $m_{A'} = 0$, corresponding to the background (see, {\it e.g.}, Appendix~A of Ref.~\cite{Asai:2021ehn}).
Therefore, the energy cut employed in the LDMX experiment with a 4~GeV electron beam can also be applied to suppress background events.
In our analysis, we impose an energy cut of $E_{\rm cut}=0.1\, E_{\rm beam}$, following the LDMX strategy (which adopts $E_{\rm cut}=0.3\, E_{\rm beam}$).
Moreover, as shown in the right panel, the recoil electron is predominantly emitted in the forward direction; however, for larger dark photon masses $m_{A'}$, the recoil-angle distribution becomes broader.

\begin{figure}[t]
\begin{center}
\includegraphics[width=9.cm]{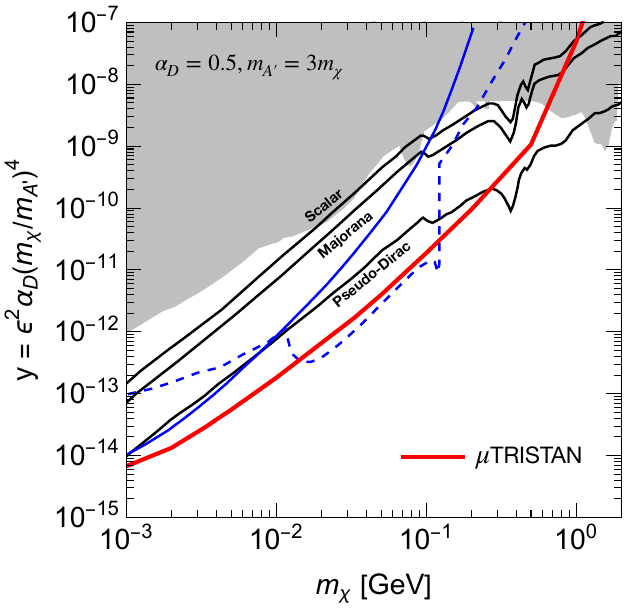}
\end{center}
\vspace{-0.4cm}
\caption{Expected 95\,\% C.L. sensitivity of a fixed-target experiment to sub-GeV dark matter models using electron/positron beams extracted from muon collider experiments.
This plot assumes $\alpha_D = 0.5$ and $m_{A'} = 3 m_{\chi}$.
The red solid curve shows the benchmark sensitivity for $E_{\mathrm{beam}}=800$ GeV ($\mu$TRISTAN), assuming $N_e=4\times10^{14}$ electrons on target selected within a momentum interval of full width $\Delta p/p=1\,\%$.
The gray regions are already excluded by other experiments, while the blue solid and dashed curves indicate the projected sensitivities of the LDMX (Phase~1) experiment to bremsstrahlung dark photon production and of the ILC-BDX experiment.
The black solid curves indicate the thermal relic targets for scalar elastic, Majorana, and pseudo-Dirac dark matter, corresponding to parameter combinations that reproduce the observed dark matter relic density.
}
\label{fig:sen_missing}
\end{figure}

We now present the sensitivity of the $\mu$TRISTAN setup to sub-GeV dark matter models in Fig.~\ref{fig:sen_missing}.
The horizontal axis represents the dark matter mass $m_{\chi}$, while the vertical axis is defined as $y=\epsilon^2 \alpha_D (m_{\chi}/m_{A'})^4$, where $\alpha_D = g_D^2/(4\pi)$.
In Fig.~\ref{fig:sen_missing}, we assume $\alpha_D=0.5$, and $m_{A'}/m_{\chi}=3$.
The red solid curve shows the 95\,\% C.L. sensitivity of the $\mu$TRISTAN setup assuming
$E_{\rm beam}=800~\mathrm{GeV}$, $N_e=4\times 10^{14}$, and an energy cut $E_{\rm cut}=0.1\,E_{\rm beam}$.
We assume a background-free scenario, for which the 95\,\% C.L. sensitivity corresponds to requiring $N_{\rm signal}\geq 3$.
The gray regions are already excluded by existing experiments\footnote{For DM-induced recoil searches, we assume pseudo-Dirac DM.} (BaBar~\cite{Berlin:2018bsc,Izaguirre:2013uxa,Essig:2013vha}, LSND~\cite{Izaguirre:2017bqb,Berlin:2018bsc,deNiverville:2011it,Berlin:2018pwi}, E137~\cite{Batell:2014mga}, MiniBooNE~\cite{MiniBooNEDM:2018cxm}, NA64~\cite{Andreev:2021fzd}, and COHERENT CsI~\cite{COHERENT:2021pvd,COHERENT:2022nrm}).
The blue solid curve corresponds to the projected sensitivity of the LDMX experiment (Phase~1) to dark-photon bremsstrahlung production, while the blue dashed curve represents the projected sensitivity of the ILC-BDX experiment~\cite{Asai:2023dzs} based on DM-induced recoil-energy searches.
The black solid curves indicate the thermal relic targets for scalar elastic, Majorana, and pseudo-Dirac dark matter~\cite{Marsicano:2018glj}, {\it i.e.}, the preferred combinations of parameters that reproduce the observed dark matter relic density.
We find that the $\mu$TRISTAN setup can probe a wide range of thermal relic targets.

Finally, we comment on the complementary role of missing-energy searches relative to recoil searches.
Unlike recoil searches, missing-energy searches have good acceptance and higher sensitivity, since the produced DM particles do not need to be directly detected. 
However, missing-energy searches can only provide indirect evidence of DM. 
Even if a missing-energy signal is observed, recoil searches are necessary to validate the existence of DM, as illustrated by the historical discovery of neutrinos.
Therefore, future recoil searches, such as ILC-BDX, remain essential despite their lower sensitivity.

\subsection{Visible-decay search in the IMCC setup}
\label{sec:visible}
The IMCC positron/electron beam is characterized by two key properties: high energy and a bunched structure.
These properties make it well suited for searches for visible decays of new light particles.
As a benchmark, we consider axion-like particles (ALPs) and scalars coupled to photons, and examine visible decay searches using the IMCC positron/electron beam below.

\subsubsection{ALP and scalar coupling to photons}\label{sec:alp}
Our main focus in the visible-decay search is on models with an axion-like particle $a$ and a scalar $\phi$ that have effective couplings to photons (see, {\it e.g.}, Refs.~\cite{Batell:2022dpx,Batell:2022xau} and references therein for reviews, with an emphasis on visible-decay searches),
\begin{align}
    \mathcal{L}\supset \frac{a}{4\Lambda_a} F_{\mu\nu}\widetilde{F}^{\mu\nu}+\frac{\phi}{4\Lambda_{\phi}}F_{\mu\nu}F^{\mu\nu},
\end{align} 
where $\widetilde{F}_{\mu\nu}=\tfrac{1}{2}\epsilon_{\mu\nu\alpha\beta}F^{\alpha\beta}$.
In particular, the ALP is a pseudo-Goldstone boson, and its mass, $m_a$, can be much smaller than the scale of its interactions with SM particles, {\it i.e.}, $m_a \ll \Lambda_a$.
The decay rate of the ALP and scalar into two photons are given by
\begin{align}
    \Gamma_{a\to 2\gamma}=\frac{m^3_a}{64\pi \Lambda^2_a},\qquad \Gamma_{\phi\to 2\gamma}= \frac{m^3_{\phi}}{64 \pi \Lambda^2_\phi}.
\end{align}
As will be discussed below, these new particles are mainly produced through interactions of photons in the electromagnetic shower, generated by the electron/positron beam, with nuclei in the fixed target (see Fig.~\ref{fig:setup_ALP}).

\subsubsection{Experimental setup}

\begin{figure}[t]
\begin{center}
\includegraphics[width=11.cm]{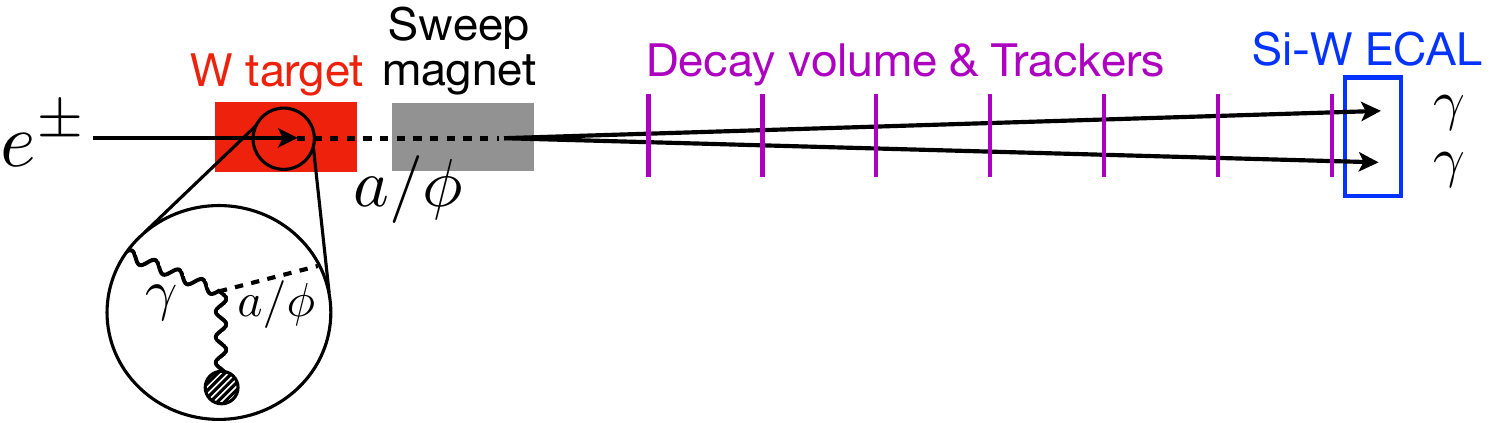}
\end{center}
\vspace{-0.4cm}
\caption{The visible-decay search setup consists of a thick tungsten target (50 cm long), a sweep magnet, a 20 m decay volume with tracking detectors, and a Si-W electromagnetic calorimeter (20\,cm $\times$ 20\,cm $\times$ 10.5 cm).
Extracted positron and electron beams from the muon collider are injected into a target,
where ALPs $a$ and scalars $\phi$ may be produced via the Primakoff process.
These particles can decay into two photons in the downstream decay volume and are detected
by a Si-W electromagnetic calorimeter.
A sweep magnet and tracking detectors are used to suppress and identify charged-particle
backgrounds.
}
\label{fig:setup_ALP}
\end{figure}

In the following analysis, we assume that the extracted electron beam carries
a fraction $r=80\,\%$ of the muon beam energy.
In the IMCC scenarios, this corresponds to an electron beam energy of
$E_{\rm beam}=4~\mathrm{TeV}$ for IMCC-1-II and IMCC-2, and
$E_{\rm beam}=1.2~\mathrm{TeV}$ for IMCC-1-I (see Table~\ref{tab:para}).
In total, we assume $N_e=10^{15}$ incoming electrons accumulated over several
years of operation, which corresponds to $10^{3}$ electrons per bunch.
We adopt the experimental setup illustrated in Fig.~\ref{fig:setup_ALP}.
The electron beam is injected into a tungsten target with a length of 50\,cm.
Downstream of the target, a magnet is installed to deflect charged particles.
This sweep magnet is not intended to completely remove high-energy charged
particles that contribute to BG events (mainly $\pi^\pm$), but rather
to reduce the number of charged particles simultaneously reaching the detector
to $\mathcal{O}(1)$ or less per bunch.
A decay volume of 20\,m is assumed downstream, in which a tracking detector is
placed to measure the trajectories of charged particles leaking from the magnet.
Further downstream, a Si-W sampling electromagnetic calorimeter (ECAL) is
installed. The transverse dimensions of the detector perpendicular to the beam
axis are set to 20\,cm $\times$ 20\,cm, and the total tungsten thickness along
the beam direction corresponds to 30 radiation lengths, equivalent to 10.5\,cm.

\subsubsection{Numerical results}

After the beam is injected into the target, in addition to charged particles such as muons and charged pions, neutral hadrons can be produced via photonuclear reactions.
Since these neutral hadrons cannot be deflected by the sweep magnet, they can constitute a potential BG.
However, as will be estimated based on PHITS-based Monte Carlo simulations below, the BG induced by neutral particles can be reduced to a negligible level by imposing suitable cuts and using a preshower detector. 
To suppress BG events and isolate the signal, we require two clusters
in a single-bunch event, each with an energy deposition exceeding 20\% of the
beam energy $E_{\rm beam}$.
For BG events, assuming $10^{3}$ electrons per bunch, the probability per bunch to satisfy this condition is $1.1\times10^{-11}$, as estimated using PHITS-based Monte Carlo simulations.
To suppress BG from charged pions, clusters induced by charged particles originating upstream of the decay volume are vetoed using tracking detectors in the decay volume together with the high-granularity Si-W calorimeter.

In addition, a preshower detector is introduced either in the front layers of the sampling calorimeter or slightly upstream of it to discriminate neutrons from photons in background events.
The total thickness of the front layers, or equivalently the tungsten converter thickness, is set to 1.35\,cm, and the presence of charged particles downstream of this region is required.
The probability that both signal photons are converted into charged particles is 90\,\%, whereas the probability that two background neutrons simultaneously interact in this region is 1.6\,\%.
We take this value as an additional reduction factor for BG events.

BG events can be further suppressed using the opening angle $\Delta\varphi$ between the photon hit positions with respect to the beam axis, as well as the reconstructed ALP mass~\cite{Ishikawa:2021qna}.
Since the reduction factor based on the reconstructed mass depends on the ALP mass, we consider only the $\Delta\varphi$-based reduction for simplicity in this paper.
For BG events, the photon hit positions are uniformly distributed over the $20~\mathrm{cm}\times20~\mathrm{cm}$ detector plane, whereas for signal events they are concentrated around $\Delta\varphi\simeq\pi$, due to the extremely small production angles of the hard bremsstrahlung photons producing the ALP.
Assuming a conservative detector position resolution of 0.93\,cm, comparable to the Moli\`ere radius of tungsten, and imposing a cut at $\Delta\varphi\simeq\pi$, the resulting reduction factor is approximately 4.6\,\%.

Consequently, when the total number of incident electrons reaches $10^{15}$, the expected number of background events is 0.008, which is sufficiently small to be neglected.
Therefore, background events are ignored in the analysis.
If further background suppression is required, it can be achieved by introducing a mass cut, reducing the number of electrons per bunch, or increasing the target length.

\begin{figure}[t]
\begin{center}
\includegraphics[width=8.cm]{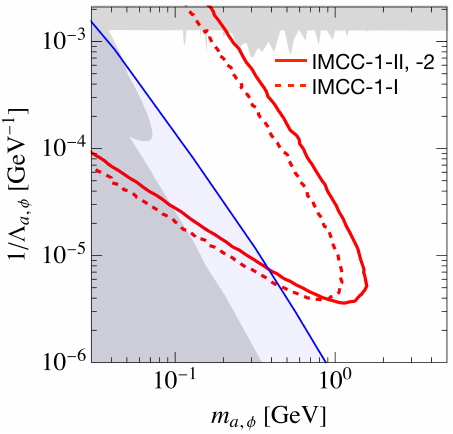}
\end{center}
\vspace{-0.4cm}
\caption{Expected 95\,\% C.L. sensitivities of a fixed-target experiment to ALPs and scalar particles using electron/positron beams extracted from muon collider experiments.
The red solid and dotted curves denote the sensitivities for $E_{\rm beam}=4~\mathrm{TeV}$ (IMCC-1-II and IMCC-2) and $1.2~\mathrm{TeV}$ (IMCC-1-I), respectively, assuming $N_e=10^{15}$ for both cases.
The gray regions are already excluded by other experiments (E137~\cite{Bjorken:1988as}, NuCal~\cite{Blumlein:1990ay}, NA64~\cite{NA64:2020qwq}, Belle II~\cite{Belle-II:2020jti}, and FASER~\cite{FASER:2024bbl}, as well as photon-beam searches~\cite{PrimEx:2010fvg,Aloni:2019ruo} and $e^+e^- \to \gamma\gamma$ measurements~\cite{Knapen:2016moh}), while the blue regions indicate the projected sensitivity of the SHiP experiment~\cite{Jerhot:2022chi}.
}
\label{fig:ALPs}
\end{figure}

Following Ref.~\cite{Sakaki:2020mqb}, we evaluate the number of signal events under the above cut (two energy depositions originating from a single electron bunch, each exceeding $20\,\%$ of the beam energy).
Figure~\ref{fig:ALPs} presents the 95\,\% C.L. sensitivity of the IMCC scenario to ALPs and scalar particles.
We assume a background-free scenario under this selection, for which the 95\,\% C.L. sensitivity corresponds
to requiring $N_{\rm signal}\geq 3$.
The red solid and dotted curves denote the sensitivities for $E_{\rm beam}=4~\mathrm{TeV}$ (IMCC-1-II and IMCC-2) and $1.2~\mathrm{TeV}$ (IMCC-1-I), respectively.
The gray regions are already excluded by other experiments (E137~\cite{Bjorken:1988as}, NuCal~\cite{Blumlein:1990ay}, NA64~\cite{NA64:2020qwq}, Belle II~\cite{Belle-II:2020jti}, and FASER~\cite{FASER:2024bbl}, as well as photon-beam searches~\cite{PrimEx:2010fvg,Aloni:2019ruo} and $e^+e^- \to \gamma\gamma$ measurements~\cite{Knapen:2016moh}), while the blue regions correspond to the projected sensitivity of the SHiP experiment~\cite{Jerhot:2022chi}.
We find that, by leveraging the extracted electron beam from muon collider experiments, short-lifetime regions of these particles can be probed.

\section{Summary}
\label{sec:sum}
Future muon collider experiments offer unique opportunities for both the discovery of new heavy particles and precision studies of the electroweak and Higgs sectors.
A distinctive feature of these experiments, not present in other accelerator facilities, is that the circulating muons are unstable and continuously produce neutrinos and electrons/positrons via their decays.
In this paper, using Monte Carlo simulations with PHITS and the tentative parameters proposed by the IMCC and $\mu$TRISTAN, we analyze the positions, angles, and energy distributions of electron and positron beams produced by muon decays in the curved sections of a muon collider ring, taking into account the effects of the bending-magnet fields (or magnets providing an equivalent magnetic field).
We find that deflections of $0.1$\text{--}$10~\mathrm{mrad}$ for the electron and positron beams can typically be obtained even for high-energy electrons with large energy fractions $\simeq 0.6$\text{--}$1.0$ of the muon beam energy, significantly larger than those achieved in the LHC extraction system.
This large deflection suggests that the extraction scheme could be practically feasible and that either the bending magnets or dedicated magnets designed for this purpose might act as a natural pre-septum magnet, potentially eliminating the need for an additional kicker.
Moreover, our analyses show that the number of extractable electrons and positrons is independent of the extraction length $L_{\rm ext}$ due to beam-transport acceptance.

We further find that, before accounting for the efficiency factor~\eqref{eq:eta}, the expected number of extractable electrons over one year of operation~\eqref{eq:hatNe} is largely independent of the experimental configuration.
In contrast, the IMCC provides a much larger number of electrons per muon bunch, whereas $\mu$TRISTAN offers a significantly higher repetition rate.
These differences indicate that the optimal applications of the extracted electron and positron beams depend strongly on the specific muon collider design.
Further details are provided in Table~\ref{tab:para}.

To highlight the physics potential of extracted electron and positron beams, we propose new searches for light particles based on such beams and evaluate their expected sensitivities.
Motivated by the distinct beam properties of the IMCC and $\mu$TRISTAN, we develop two complementary search strategies: a missing energy/momentum search in the $\mu$TRISTAN setup and a visible-decay search in the IMCC setup.
The corresponding experimental configurations are illustrated in Figs.~\ref{fig:ext}, \ref{fig:LDMX} and \ref{fig:setup_ALP}.
In the missing energy/momentum search, we consider sub-GeV dark matter interacting with the SM via a hidden-photon mediator, while in the visible-decay search we study axion-like particles and light scalars coupled to photons.
In both cases, we demonstrate that the extracted beams can probe parameter regions beyond current experimental bounds and projected future sensitivities.
Taken together, these search strategies can be pursued in parallel with the primary physics program at the main interaction point, significantly broadening the muon collider's reach for physics beyond the SM.

Finally, we briefly discuss the practical feasibility of implementing beam extraction.
Our analysis relies exclusively on the geometrical size and magnetic field strength along the curved sections of the muon collider ring, without considering the detailed design of the muon collider.
Incorporating such detailed facility designs may introduce additional constraints on the extraction of electron and positron beams.
We expect that this work provides a useful basis for future studies of the practical implementation of electron and positron beam extraction at muon collider experiments.

\appendix

\section{Irreducible backgrounds in the missing search}
\label{app:Irre}
\subsection{Neutrino trident production}
The missing signature with a single positron or electron track can originate from coherent neutrino trident production, in which a positron or electron scatters off a nucleus N, producing a pair of neutrinos in the final state, $
    e^{\pm} {\rm N} \to e^{\pm}\nu_e \bar{\nu}_e  {\rm N}$.
According to the Equivalent Photon Approximation (EPA), the approximate total coherent cross section producing the final-state particles $e^\pm \nu_e \bar{\nu}_e$, $\sigma(e^\pm  {\rm N} \to e^\pm \nu_e \bar{\nu}_e {\rm N} )$, can be estimated by computing the cross section for the scattering of the incoming particle with a real photon, integrated over the energy spectrum of photons~\cite{Altmannshofer:2014pba,Ballett:2018uuc},
\begin{align}
    \sigma (e^\pm {\rm N} \to e^\pm \nu_e \bar{\nu}_e {\rm N} ) \simeq \int d P(Q^2,\hat{s}) \sigma\left(\gamma e^\pm  \to e^\pm \nu_e \bar{\nu}_e ;\hat{s},Q^2 =0\right)\cdot \Theta \left(E_{\rm cut}-E_{\rm recoil}\right), \label{eq:coh_cross}
\end{align}
where the cross section for the process $ \gamma e^\pm \to e^\pm \nu_e \bar{\nu}_e $, denoted as
$\sigma\left(\gamma e^\pm \to e^\pm \nu_e \bar{\nu}_e ;\hat{s},Q^2=0\right)$, depends on the center-of-mass energy of the N-photon system, $\sqrt{\hat{s}}$.
Here, $dP(Q^2, \hat{s})$ corresponds to the energy spectrum of the virtual photons; that is, the probability of emitting a virtual photon with four-momentum transfer $Q^2$, resulting in a center-of-mass energy $\sqrt{\hat{s}}$.
For $Q^2 = 0$, the center-of-mass energy is given by $\hat{s} \simeq 2 E_{\rm beam} E_{\gamma} (1 - \cos\theta_{e\gamma})$.
For trident scattering off a nuclus target, this probability can be approximated by~\cite{Altmannshofer:2014pba,Belusevic:1987cw,Ballett:2018uuc}
\begin{align}
    d P(Q^2,\hat{s}) = \frac{Z^2 e^2}{4\pi^2}\left|F(Q^2)\right|^2
    \frac{d \hat{s}}{\hat{s}} \frac{d Q^2}{Q^2},
\end{align}
where an analytic expression for the form factor in the coherent scattering regime is given by
\begin{align}
    F\left(Q^2\right)=\frac{3\pi a}{r_0^2+\pi^2 a^2}
    \frac{\pi a \coth \left(\pi Q a\right)\sin \left(Qr_0\right)-r_0 \cos \left(Q r_0\right)}{Q r_0 \sinh \left(\pi Q a\right)},
\end{align}
with $r_0=1.126 A^{1/3}$ fm and $a=0.523$ fm.
The kinematically allowed region in the $(Q^2, \hat{s})$ plane can be expressed as
\begin{align}
    &Q^2_{\rm min} = \frac{M_{{\rm N}}\hat{s}^2}{2E_{\rm beam}\left(2 E_{\rm beam} M_{{\rm N}}-\hat{s}\right)},~~~~Q^2_{\rm max}=\hat{s},
    \\
    &\hat{s}_{\rm min} =\frac{E_{\rm beam}}{2 E_{\rm beam} +M_{{\rm N}}}\left(2 E_{\rm beam} M_{{\rm N}} 
    -\Delta
    \right)=0,
    \\
    &\hat{s}_{\rm max}=\frac{E_{\rm beam}}{2E_{\rm beam} +M_{{\rm N}}}
    \left(
    2E_{\rm beam} M_{{\rm N}}
    +\Delta
    \right)=\frac{4E_{\rm beam}^2M_{{\rm N}}}{2E_{\rm beam} +M_{{\rm N}}},
\end{align}
with a nuclear mass $M_{\rm N}$ and $\Delta = 2 E_{\rm beam} M_{\rm N}$.
From these, the number event of the process per incident positron/electron is given as
\begin{align}
    P_{e^\pm {\rm N}\to e^\pm \nu_e \bar{\nu}_e {\rm N}}= n_{{\rm N}}\cdot l_{\rm target}\cdot \sigma\left(e^\pm {\rm N}\to e^\pm \nu_e \bar{\nu}_e {\rm N} \right).\label{eq:tri_rate}
\end{align}
Combining Eqs.~\eqref{eq:coh_cross} and \eqref{eq:tri_rate} with the photo-production cross section for the process $\gamma e^\pm \to e^\pm \nu_e \bar{\nu}_e $, computed using \texttt{MadGraph5\_aMC@NLO}~\cite{Alwall:2011uj}, we find that, for $E_{\rm beam}=800$ GeV, the probability is approximately $P_{e^\pm {\rm N} \to e^\pm \nu_e \bar{\nu}_e {\rm N}} \sim 10^{-16}$ without any energy cut, and remains $\sim 10^{-16}$ even for $E_{\rm cut} = 0.1\, E_{\rm beam}$, and $0.01\, E_{\rm beam}$.

\subsection{M{\o}ller scattering + charged-current quasi-elastic reaction}
A missing energy signature with a single electron track can arise from charged current quasi-elastic (CCQE) interactions, $e^- p \to n \nu_e$, initiated by an electron that has undergone M{\o}ller scattering ($e^- e^- \to e^- e^-$) in the thin target.
The differential cross section for M{\o}ller scattering is given in Refs.~\cite{Izaguirre:2014bca,Kahn:2018cqs}:
\begin{align}
    \frac{d \sigma_{e^{-} e^-\to e^{-} e^-}}{dK_{\rm recoil}}=\frac{2\pi \alpha^2}{m_e K^2_{\rm recoil}}
    \left(
    1-\frac{K_{\rm recoil}}{E_{\rm beam}}
    +
    \frac{K_{\rm recoil}^2}{2E_{\rm beam}^2}
    \right),
\end{align}
where $E_{\rm beam}$ and $K_{\rm recoil}$ denote the incoming and recoil electron energies, respectively, evaluated in the lab frame.
We denote the initial electron four-momenta by $p_{\rm beam}$ and $p_{\rm rest}$, and the final electron four-momenta by $p_{\rm out}$ and $p_{\rm recoil}$ as follows:
\begin{align}
    p_{\rm beam}=(E_{\rm beam},\, \vec{p}_{\rm beam}),\quad p_{\rm rest}=(m_e,\,0),\quad p_{\rm out}=(E_{\rm out},\,\vec{p}_{\rm out}),\quad p_{\rm recoil}=(E_{\rm recoil},\, \vec{p}_{\rm recoil}),
\end{align}
where $E_{\rm recoil}=m_e+K_{\rm recoil}$.
From energy conservation, $E_{\rm out}=E_{\rm beam}+m_e -E_{\rm recoil}$, and from three-momentum conservation, $\vec{p}_{\rm out}^2=\vec{p}_{\rm beam}^2+\vec{p}^2_{\rm recoil}-2|\vec{p}_{\rm beam}||\vec{p}_{\rm recoil}|\cos\theta$, we obtain the relation $f(E_{\rm rec})=\cos\theta$ for a given $E_{\rm beam}$, where
\begin{align}
    f(E_{\rm recoil}):=-\frac{1}{2|\vec{p}_{\rm beam}||\vec{p}_{\rm recoil}|}\left[\left(E_{\rm beam}+m_e-E_{\rm recoil}\right)^2-m_e^2 -\left(E_{\rm beam}^2-m_e^2\right)-(E_{\rm recoil}^2-m_e^2)\right].
\end{align}
Namely, $df(E_{\rm recoil})/dE_{\rm recoil}\times dE_{\rm recoil}/d\theta=-\sin\theta$.
Hence the recoil energy $E_{\rm recoil}$ is extremized at $\theta = 0$ and $\pi$.
At these angles, the condition $f(E_{\rm recoil})=\cos\theta$ leads to
\begin{align}
    \left[\left(E_{\rm beam}+m_e-E_{\rm recoil}\right)^2-m_e^2 -\left(E_{\rm beam}^2-m_e^2\right)-(E_{\rm recoil}^2-m_e^2)\right]^2=4\left(E_{\rm beam}^2-m_e^2\right)\left(E_{\rm recoil}^2-m_e^2\right).
\end{align}
Solving this equation, we find $E_{\rm recoil}^{(\rm max)}=E_{\rm beam}$, and $E_{\rm recoil}^{(\rm min)}=m_e$.

For an infinitesimal thickness $dz$, the probability for a CCQE reaction $e^- p \to n \nu_e$ to occur after a M{\o}ller scattering is given by~\cite{Kahn:2018cqs}
\begin{align}
    d P_{e^{-} e^- \to e^{-} e^-+{\rm CCQE}}= dz \cdot n_{e^-} \int_{0}^{E_{\rm beam}-m_e}d K_{\rm recoil} \frac{d \sigma_{e^{-} e^-\to e^{-} e^-}}{dK_{\rm recoil}}\cdot P_{\rm CCQE} \cdot \Theta \left(E_{\rm cut}- (E_{\rm beam}-K_{\rm recoil})\right),
\end{align}
where $E_{\rm cut}$ is the energy cut applied to the outgoing single-electron track, and 
    \begin{align}
    P_{\rm CCQE}=n_p \sigma_F (l_{\rm target}-z),\quad \sigma_F=\frac{G_F^2 m_p K_{\rm recoil}}{2\pi},
    \end{align}
    denotes the probability of a CCQE event for an electron with energy $K_{\rm recoil}$, after it undergoes M{\o}ller scattering at position $z \in (0,, l_{\rm target})$, and then the recoiled electron traverses the remaining length $(l_{\rm target}-z)$ of the target with proton density $n_p = n_{e^-} = 4.7\times 10^{24}~\text{cm}^{-3}$, within the one-dimensional approximation.
    For $E_{\rm beam}=800$ GeV, the probability is of order $10^{-21}$ for $E_{\rm cut}=0.1\,E_{\rm beam}$, while it decreases to $10^{-22}$ for $E_{\rm cut}=0.01\,E_{\rm beam}$.

\clearpage
\bibliography{MC.bib}

\end{document}